\documentclass[twocolumn,showpacs,amsmath,amssymb,superscriptaddress]{revtex4-1}
\usepackage{dcolumn,braket,color,amsmath,footnote,hyperref,bm}
\usepackage{graphicx,bm,url,longtable}
\usepackage{tabularx,multirow,braket}
\usepackage{fancybox}
\usepackage{tikz}
\usetikzlibrary{matrix}

\newcommand\+{\dagger}

\begin{document}

\title{Shape phase transitions in odd-A Zr isotopes}

\author{K.~Nomura}
\email{knomura@phy.hr}
\affiliation{Department of Physics, Faculty of Science, University of Zagreb, HR-10000, Croatia}

\author{T.~Nik\v si\'c}
\affiliation{Department of Physics, Faculty of Science, University of Zagreb, HR-10000, Croatia}

\author{D.~Vretenar}
\affiliation{Department of Physics, Faculty of Science, University of Zagreb, HR-10000, Croatia}

\date{\today}

\begin{abstract}
Spectroscopic properties that characterize shape phase transitions 
in neutron-rich odd-A Zr isotopes 
are investigated using the framework of nuclear 
density functional theory and particle-core coupling. 
The interacting-boson Hamiltonian of the 
even-even core nuclei, and the single-particle 
energies and occupation probabilities of the unpaired 
neutron are completely determined by deformation 
constrained self-consistent mean-field calculations based on the 
relativistic Hartree-Bogoliubov model with 
a choice of a universal energy density 
functional and pairing interaction. 
The triaxial $(\beta,\gamma)$ deformation energy surfaces 
for even-even $^{94-102}$Zr indicate the occurrence of a transition 
from triaxial or $\gamma$-soft ($^{94,96}$Zr) to prolate 
($^{98}$Zr), and  triaxial  ($^{100,102}$Zr) shapes. 
%The resultant energy spectra for even-even Zr 
%show a phase transitional, rapid decrease of the low-lying levels 
%from $^{96}$Zr to $^{100}$Zr. 
The corresponding low-energy excitation spectra of the odd-A Zr isotopes 
are in very good agreement with recent experimental results. 
Consistent with the structural evolution of the neighboring 
even-even Zr nuclei, the state-dependent effective deformations and 
their fluctuations in the odd-A isotopes indicate a pronounced discontinuity 
around the transitional nucleus $^{99}$Zr. 
\end{abstract}

\maketitle

\section{Introduction}

For many years the structure of neutron-rich nuclei with mass number $A\approx 100$ has been a 
challenging topic for experiments that use radioactive-ion beams. This particular mass region has also attracted considerable attention in theoretical studies due to its rich microscopic structure. The effective interaction between 
nucleons determines the corresponding shell structure and 
gives rise to various shapes, quantum (shape) phase transitions 
\cite{cejnar2010}, and shape coexistence \cite{heyde2011}.  
Since neutron-rich nuclei in this mass region are also involved in the rapid neutron-capture 
process, an accurate theoretical description of their low-lying structure and 
transition rates is important for modelling the formation of chemical 
elements in various astrophysical scenarios. In many cases the low-energy 
structure is so rich that it provides an ideal  
testing ground for theoretical models.

Recently a number of experimental and theoretical studies of spectroscopic  
properties of even-even Zr isotopes have been reported. 
Most experimental results have suggested the occurrence of 
shape coexistence in $^{96}$Zr \cite{sazonov2019} and 
$^{98}$Zr \cite{singh2018,witt2018}, 
a quantum phase transition around the neutron 
number $N\approx 60$ \cite{kremer2016,ansari2017}, 
and $\gamma$-soft and triaxial shapes at $^{100,102}$Zr \cite{urban2019}. 
Theoretical studies have generally confirmed 
these experimental findings 
\cite{togashi2016,nomura2016zr,sazonov2019,garciaramos2019,gavrielov2019}. 
In contrast, much less theoretical research 
has been devoted to shape-phase transitions in odd-A Zr nuclei, for which  
in the last couple of years several measurements of various spectroscopic 
properties have been reported, e.g., 
$^{97}$Zr \cite{rzacaurban2018}, and $^{99}$Zr 
\cite{spagnoletti2019,boulay2020}.

A microscopic calculation of spectroscopic properties of odd-mass nuclei 
is a challenging task, because in odd-A systems  
one has to take explicitly into account both single-particle and collective 
degrees of freedom \cite{BM}. 
We have developed a theoretical method \cite{nomura2016odd} for 
computing spectroscopic properties of odd-A nuclei, based on the framework of 
nuclear density functional theory (DFT) \cite{bender2003,vretenar2005,robledo2019}
and the particle-core coupling scheme \cite{BM}. 
In this approach the even-even core is described with the 
interacting boson model (IBM) \cite{IBM}, and 
the particle-core coupling is fashioned using 
the interacting boson-fermion model (IBFM) \cite{IBFM}. 
In a first step a set of constrained self-consistent mean-field (SCMF) calculations  
is performed for each even-even mass nucleus to provide the potential energy surface (PES). 
By mapping the SCMF energy surface 
onto the expectation value of the IBM Hamiltonian, 
the parameters of the interaction terms of the even-even (boson) core 
Hamiltonian are completely determined. 
The same SCMF calculations also provide the spherical 
single-particle energies and occupation probabilities for the 
odd nucleon, and these quantities are used as input to 
construct the boson-fermion interactions. 
Even though a few boson-fermion interaction strengths 
have to be adjusted to the empirical low-energy spectra 
for each odd-A nucleus, the method has allowed for 
a systematic, detailed, and computationally efficient 
description of spectroscopic properties of nuclei with 
odd nucleon number(s). 
So far, this method has been applied to a variety of 
nuclear structure phenomena in odd-mass and odd-odd nuclei, 
including quantum phase transitions in axially-symmetric 
\cite{nomura2016qpt} and $\gamma$-soft \cite{nomura2017odd-1} 
odd-A nuclei, octupole correlations in 
neutron-rich Ba isotopes \cite{nomura2018oct}, 
chiral band structure in the mass $A\approx 130$ \cite{nomura2020cs}
region, and $\beta$-decay \cite{nomura2020beta-1,nomura2020beta-2}.

The scope of this work is a simultaneous description of quantum phase transitions 
that are supposed to take place 
in the even-even and odd-A Zr isotopes, using the aforementioned 
theoretical method. 
Here we consider the even-even isotopes $^{94-102}$Zr and 
the neighbouring odd-neutron nuclei $^{95-103}$Zr. 
The underlying SCMF calculations are carried out within the 
framework of the relativistic Hartree-Bogoliubov method 
with the density-dependent point-coupling (DD-PC1) \cite{DDPC1} 
energy density functional and a separable pairing force \cite{tian2009}. 
SCMF calculations based on the DD-PC1 functional have been 
successfully applied to various static and dynamic 
properties of finite nuclei, such as the phenomena of 
quantum phase transitions \cite{nomura2014,nomura2016qpt,nomura2017odd-1}, 
triaxial deformations \cite{nomura2012tri,niksic2014}, octupole correlations 
\cite{nomura2014,nomura2018oct}, shape coexistence \cite{li2016}, 
clustering \cite{marevic2018}, and fission dynamics \cite{zhao2016,zhao2019b}.

The article is organized as follows. In Sec.~\ref{sec:scmf} 
the SCMF energy surfaces for the even-even Zr isotopes 
are discussed. Section~\ref{sec:ibfm} illustrates 
the procedure to construct the bosonic 
and particle-core (IBFM) Hamiltonians for the even-even and 
odd-A Zr isotopes based on the DFT SCMF calculations. 
In Sec.~\ref{sec:results} we discuss spectroscopic properties 
of even-even and odd-A Zr isotopes in comparison to available data, including low-energy 
excitation spectra and electromagnetic transition rates, as well as  
possible signatures of quantum phase transitions (Sec.~\ref{sec:qpt}).
Section~\ref{sec:summary} contains a brief summary of the principal results.

\begin{figure*}[htb!]
\begin{center}
\includegraphics[width=0.7\linewidth]{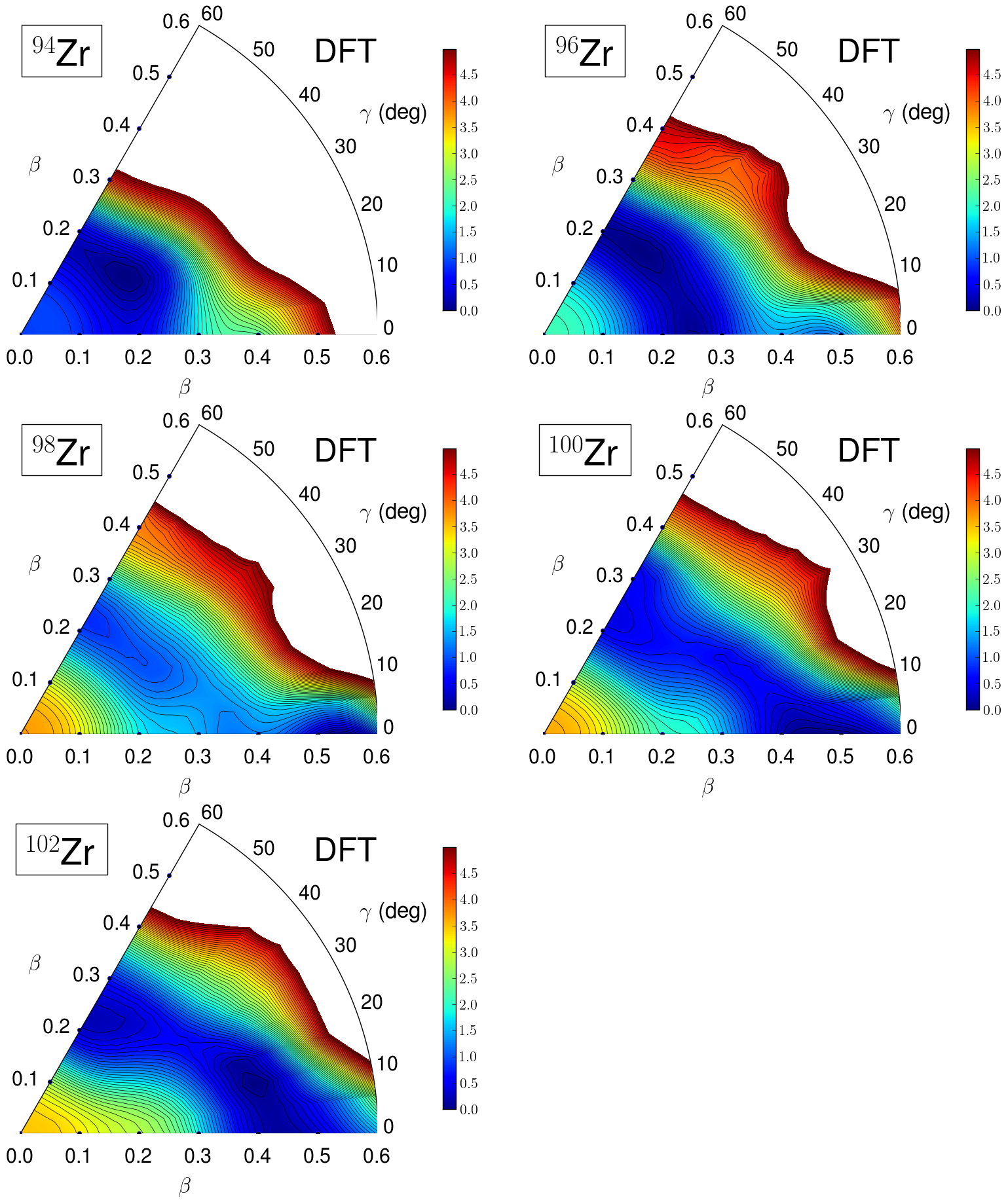}
\caption{(Color online) SCMF $(\beta,\gamma)$ deformation energy
 surfaces (in MeV) for the even-even nuclei $^{94-102}$Zr, 
 obtained from constrained relativistic Hartree-Bogoliubov 
 calculations using the DD-PC1 functional \cite{DDPC1} 
 and a separable pairing force \cite{tian2009}. The total SCMF 
 energies are plotted up to 5 MeV with respect to the 
 global minimum. The energy difference 
 between neighbouring contours is 100 keV. } 
\label{fig:pes-dft}
\end{center}
\end{figure*}

\section{Self-consistent mean-field energy surfaces for even-even Zr isotopes \label{sec:scmf}}

The first step of the analysis is a set of constrained SCMF 
calculations of potential energy surfaces for the 
even-even core nuclei, performed using 
the relativistic Hartree-Bogoliubov method \cite{vretenar2005} 
with the density-dependent 
point coupling (DD-PC1)  \cite{DDPC1} functional 
for the particle-hole channel, and a separable 
pairing force of finite range \cite{tian2009} 
in the particle-particle channel. 
The constraints imposed in the SCMF calculations 
are the mass quadrupole moments, 
which are represented by the dimensionless 
quadrupole deformation parameters 
$\beta$ and $\gamma$ \cite{BM}. 

In Fig.~\ref{fig:pes-dft} we display the SCMF $(\beta,\gamma)$ 
energy surfaces for $^{94-102}$Zr. 
Several remarkable  
features appear already at the mean-field level. 
The nucleus $^{94}$Zr exhibits a 
pronounced triaxial minimum at 
$\gamma\approx 40^\circ$, even though it is located near 
the neutron shell closure at $N=50$. 
For $^{96}$Zr, the potential becomes more $\gamma$ soft, 
and essentially two shallow minima appear, one on the prolate 
and the other on the oblate side. 
A prolate local minimum between $\beta=0.4$ and 0.5 
is also visible. 
The structure appears to change significantly at $^{98}$Zr: 
while the surface is still rather flat in the $\gamma$ direction for the 
interval $0.2\leqslant\beta\leqslant 0.3$, a pronounced 
prolate minimum develops at around $\beta=0.5$ and 
becomes the equilibrium configuration. 
This prolate minimum develops even further for $^{100}$Zr but, 
compared to $^{98}$Zr, the surface again becomes softer in $\gamma$. 
Finally, in $^{102}$Zr a triaxial global minimum is found 
at $\gamma\approx 15^\circ$. 
Those $\gamma$-soft and triaxial shapes obtained for  
$^{100,102}$Zr are compatible with recent experimental 
results \cite{urban2019}.

It might be useful to note some predictions obtained using different EDFs. 
Especially, results of Hartree-Fock-Bogoliubov calculations 
based on the Gogny-D1S \cite{D1S} EDF are available \cite{CEA}. 
The Gogny-HFB calculations predict an almost spherical 
shape for $^{94}$Zr, and a weakly-deformed oblate shape 
for $^{96}$Zr. They also determine a coexistence of oblate 
(at $\beta\approx 0.2$) and prolate (at $\beta\approx 0.5$) 
minima in $^{96}$Zr consistent with the result of the present work, but in the  
former case the global minimum is on the oblate side. For the deformed 
nuclei $^{100,102}$Zr, the Gogny-HFB surfaces appear  
rather similar to the present results. 
The Gogny-HFB calculation with the D1M EDF \cite{D1M} 
has also been reported in Ref.~\cite{nomura2016zr}. 
The D1M energy surfaces are generally 
softer, but not strikingly different from the D1S ones. A noticeable difference 
between the two Gogny EDFs 
is that with the D1M EDF an oblate global minimum is 
obtained for $^{100}$Zr. 

%-----------------------------------------------------------------------
%
% 	MAPPED IBM-2 ENERGY SURFACE
%
%-----------------------------------------------------------------------
\begin{figure*}[htb!]
\begin{center}
\includegraphics[width=0.7\linewidth]{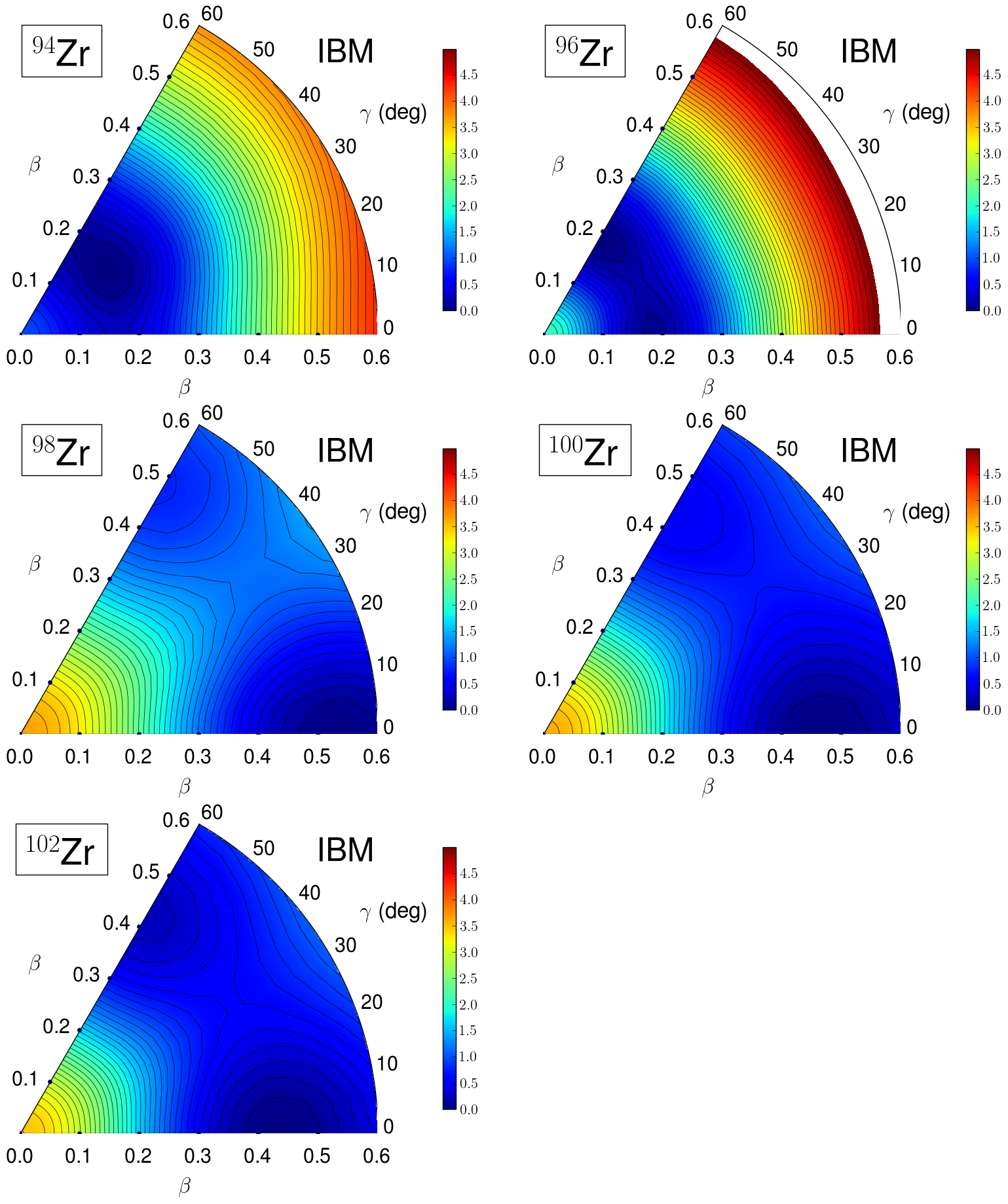}
\caption{(Color online) The bosonic energy surfaces based on the 
IBM-2 Hamiltonian in Eq.~(\ref{eq:ibm2}), with the parameters determined 
by the corresponding constrained SCMF calculations.} 
\label{fig:pes-ibm}
\end{center}
\end{figure*}

\section{Construction of the fermion-boson Hamiltonian\label{sec:ibfm}}

To calculate spectroscopic properties of nuclei, 
the static mean-field method has to be extended to include 
collective correlations that arise from symmetry restoration 
and fluctuations around  mean-field minima \cite{RS}. 
In the present work collective correlations are taken into 
account by mapping the SCMF solutions 
onto the corresponding interacting-boson systems \cite{nomura2008}. 
The coupling of the odd nucleon to the even-even core is described within the 
neutron-proton interacting boson-fermion 
model (denoted hereafter as IBFM-2). 

The complete IBFM-2 Hamiltonian consists of the 
neutron-proton IBM (IBM-2) \cite{otsuka1978} 
Hamiltonian $\hat H_{\mathrm B}$ 
for the even-even core nucleus, the 
single-neutron or proton Hamiltonian $\hat H_\text{F}^\rho$ 
$(\rho=\nu/\pi)$, and the Hamiltonian that represents the coupling 
between the odd neutron/proton and the boson core $\hat H_\text{BF}^{\rho}$: 
\begin{align}
\label{eq:ham}
 \hat H_\text{} = \hat H_\text{B} + \hat H_\text{F}^\nu + \hat
 H_\text{F}^\pi + \hat
  H_\text{BF}^\nu + H_\text{BF}^\pi. 
\end{align}
For the IBM-2 Hamiltonian we employ the following form, which has 
been shown \cite{nomura2016zr} to provide a good description
of spectroscopic data in this mass regions:
\begin{align}
\label{eq:ibm2}
 \hat H_{\text{B}} = \epsilon(\hat n_{d_\nu} + \hat n_{d_\pi})+\kappa\hat
  Q\cdot\hat Q  + 
  \kappa'\sum_{\rho^{\prime}\neq\rho}\hat T_{\rho\rho\rho^{\prime}} 
  + \kappa''\hat L\cdot\hat L,
\end{align}
where the first term $\hat n_d=\hat n_{d_\nu} + \hat n_{d_\pi}$, with $\hat
n_{d_\rho}=d^{\dagger}_{\rho}\cdot\tilde d_{\rho}$ ($\rho=\nu,\pi$), 
represents the $d$-boson number operator, and 
$\hat Q = \hat Q_{\nu} + \hat Q_{\pi}$ is the quadrupole operator with 
$\hat Q_{\rho}=s^{\dagger}_{\rho}\tilde
d_{\rho}+d^{\dagger}_{\rho}\tilde
s_{\rho}+\chi_{\rho}[d^{\dagger}_{\rho}\times\tilde d_{\rho}]^{(2)}$. 
The third term is a specific three-body boson 
interaction \cite{nomura2012tri} with $\hat T_{\rho\rho\rho^{\prime}} =
\sum_{L}[d^{\dagger}_{\rho}\times d^{\dagger}_{\rho}\times
d^{\dagger}_{\rho^{\prime}}]^{(L)}\cdot [\tilde d_{\rho^{\prime}}\times
\tilde d_{\rho}\times\tilde d_{\rho}]^{(L)}$, where $L$ denotes the
total angular momentum of the boson system. 
As in Refs.~\cite{nomura2012tri,nomura2016zr}, we consider only 
the $L=3$ terms, since they play a dominant role in producing 
minima at $\gamma\approx 30^\circ$. 
The last term in Eq.~(\ref{eq:ibm2}) is the rotational Hamiltonian  
with the angular momentum operator 
$\hat L=\hat L_\nu + \hat L_\pi = \sqrt{10}\sum_{\rho=\nu,\pi}[d^\+_{\rho}\times \tilde d_{\rho}]^{(1)}$.

The single-nucleon Hamiltonian in Eq.(\ref{eq:ham}) reads: 
\begin{align}
\label{eq:ham-f}
 \hat H_\text{F}^\rho = -\sum_{j_\rho}\epsilon_{j_\rho}\sqrt{2j_\rho+1}
  (a_{j_\rho}^\dagger\times\tilde a_{j_\rho})^{(0)}
\end{align}
with $\epsilon_{j_\rho}$  the single-particle energy
of the spherical orbital ${j_\rho}$. 
For the boson-fermion interaction $\hat H_\mathrm{BF}^\rho$, 
we employ the commonly used form \cite{IBFM}: 
\begin{align}
\label{eq:ham-bf}
 \hat H_\text{BF}^\rho = \Gamma_\rho\hat Q_{\rho'}\cdot\hat q_{\rho} 
+
  \Lambda_\rho\hat V_{\rho'\rho} + A_\rho\hat n_{d_{\rho}}\hat n_{\rho}
\end{align}
where $\rho'\neq\rho$. 
The first, second, and third term in the equation above are the
quadrupole dynamical, exchange, and monopole interactions, respectively.
It is assumed that both the dynamical and exchange terms 
are dominated by the interaction between unlike particles (i.e., between
the odd neutron and the proton bosons or between the odd proton and the neutron
bosons), and that, for the monopole term, the interaction between
like-particles (i.e., between the odd neutron and the neutron bosons or between
the odd proton and the proton bosons) plays a dominant role \cite{IBFM}.
The fermionic quadrupole operator $\hat q_\rho$ reads: 
\begin{align}
\hat q_\rho=\sum_{j_\rho j'_\rho}\gamma_{j_\rho j'_\rho}
(a^\dagger_{j_\rho}\times\tilde a_{j'_\rho})^{(2)},
\end{align} 
where $\gamma_{j_\rho j'_\rho}=(u_{j_\rho}u_{j'_\rho}-v_{j_\rho}v_{j'_\rho})
Q_{j_\rho j'_\rho}$ and  $Q_{j_\rho j'_\rho}=\langle
l\frac{1}{2}j_{\rho}||Y^{(2)}||l'\frac{1}{2}j'_{\rho}\rangle$. 
The exchange term $\hat V_{\rho'\rho}$ in Eq.~(\ref{eq:ham-bf}) can be written as: 
\begin{align}
\label{eq:exc}
 \hat V_{\rho'\rho} 
 &= 
 -(s_{\rho'}^\dagger\tilde d_{\rho'})^{(2)}
\cdot
\Bigg\{
\sum_{j_{\rho}j'_{\rho}j''_{\rho}}
\sqrt{\frac{10}{N_\rho(2j_{\rho}+1)}}\beta_{j_{\rho}j'_{\rho}}\beta_{j''_{\rho}j_{\rho}}
\nonumber \\
&:((d_{\rho}^\dagger\times\tilde a_{j''_\rho})^{(j_\rho)}\times
(a_{j'_\rho}^\dagger\times\tilde s_\rho)^{(j'_\rho)})^{(2)}:
\Bigg\} + (H.c.), \nonumber \\
\end{align}
with $\beta_{j_{\rho}j'_{\rho}}=(u_{j_{\rho}}v_{j'_{\rho}}+v_{j_{\rho}}u_{j'_{\rho}})Q_{j_{\rho}j'_{\rho}}$.

In this work, the nearest doubly-magic nucleus $^{100}$Sn 
is taken as the boson vacuum. 
The neutron boson number $N_\nu$ is then 
equal to the number of valence neutron pairs, that is, 
$N_\nu=2,3,4,5$ and 6 for the even-even nuclei from 
$^{94}$Zr to $^{102}$Zr, respectively.
The proton boson number $N_\pi=5$ 
is fixed and equals the number of proton hole pairs. 
We note that in several previous IBM calculations 
\cite{gavrielov2019,garciaramos2019} the proton $Z=40$ 
sub-shell was taken as the inert core for the proton bosons in Zr isotopes. 
In those studies two independent  IBM Hamiltonians, 
one for the regular configuration with the proton boson number $N_\pi=0$,  
and the other for the intruder configuration with $N_\pi=2$ associated with the proton 
two-particle-two-hole excitation across the shell $Z=40$, are 
considered and allowed to mix in order to account for shape coexistence \cite{duval1981}. 
It is, however, beyond the scope of the present work to include 
intruder configurations and the corresponding configuration mixing. 
The $Z=40$ sub-shell could also be used here 
as the proton inert core but, from a practical point of view, the IBM model space with the 
proton boson number $N_\pi=0$ plus the neutron boson number 
$2\leqslant N_\nu\leqslant 6$ would be far too small for a quantitative description of 
collective physical observables. In addition, for $N_\pi=0$ the dynamical and exchange 
odd neutron -- boson interactions in Eq.~(\ref{eq:ham-bf}) do not contribute to odd-A Zr isotopes. 

The structure of the odd-A Zr nuclei is described as a system with 
a single (unpaired) neutron coupled to the even-even boson-core 
with mass number $A-1$. 
For the fermion valence space, we consider the full neutron major 
shell $N=50-82$, i.e., the $3s_{1/2}$, $2d_{3/2}$, $2d_{5/2}$, and $1g_{7/2}$ 
spherical orbitals for positive-parity states, and the unique-parity 
$1h_{11/2}$ orbital for negative-parity states.

The first step in the construction of the particle-boson Hamiltonian 
Eq.~(\ref{eq:ham}) is to specify the strength parameters 
for the IBM-2 Hamiltonian $\hat H_\mathrm{B}$. 
The parameters 
$\epsilon$, $\kappa$, $\chi_\nu$, $\chi_\pi$, and $\kappa'$ 
are completely determined by mapping the SCMF 
energy surface in the vicinity of the global minimum 
onto the expectation value of the IBM-2 Hamiltonian 
in the boson coherent state \cite{nomura2008}, i.e., 
$E_{\mathrm{SCMF}}(\beta,\gamma)\approx E_{\mathrm{IBM}}(\beta,\gamma)$. 
Only the strength parameter $\kappa''$ of the $\hat L\cdot L$ term 
has been determined separately, 
in such a way \cite{nomura2011rot} that the cranking moment 
of inertia in the bosonic intrinsic state should reproduce
the one computed by the SCMF within the relevant range 
of $|\beta|\leq 0.6$. 
The mapped IBM-2 energy surfaces, 
depicted in Fig.~\ref{fig:pes-ibm}, reproduce the 
corresponding SCMF surfaces. 
In addition, we list in Table~\ref{tab:ibm2para} the strength 
parameters for the boson-core Hamiltonian. The positive sign 
of the parameter $\kappa'$ for $^{94}$Zr leads to a   
triaxial minimum, while the opposite sign 
obtained for all the other nuclei produces the 
two minima on the energy surface corresponding to prolate and oblate shapes. 
A previous IBM calculation of Ref.~\cite{leviatan2016} has also used 
the three-body term to produce the two minima. 
However, the three-body term 
has a rather minor effect on the excitation spectra 
except for the $\gamma$ band \cite{nomura2012tri}, 
and its contribution is shown to be even weaker 
when the strength parameter $\kappa'$ has a negative sign. 
It is, therefore, expected that the contribution of   
this term to the low-lying states in the odd-A systems, 
at least near the yrast line, is also small. 
In addition, since the current IBFM code is limited 
to two-body boson interactions, 
in the following calculations for the odd-A Zr isotopes 
the three-body boson terms are not included. 

%-----------------------------------------------------------------------
%
%	TABLE OF IBM PARAMETERS
%
%-----------------------------------------------------------------------
\begin{table}[htb!]
 \begin{center}
\caption{\label{tab:ibm2para} 
Strength parameters of 
the IBM-2 Hamiltonian $\hat H_\mathrm{B}$ for the even-even 
nuclei $^{94-102}$Zr. All the parameters, 
except the dimensionless $\chi_\nu$ and $\chi_\pi$, 
are in units of MeV.}
\begin{ruledtabular}
  \begin{tabular}{ccccccc}
   & $\epsilon$  & $\kappa$ & $\chi_\nu$ & $\chi_\pi$ & $\kappa'$ & $\kappa''$ \\ 
\hline
$^{94}$Zr & 0.501 & $-$0.075 & $-$0.06 & 0.21 & 0.28 & 0.029 \\
$^{96}$Zr & 0.345 & $-$0.090 & $-$0.35 & 0.24 & $-$0.12 & 0.051  \\
$^{98}$Zr & 0.284 & $-$0.073 & $-$0.54 & 0.11 & $-$0.32 & 0.032  \\
$^{100}$Zr & 0.036 & $-$0.047 & $-$0.45 & 0.20 & $-$0.12 & 0.002 \\
$^{102}$Zr & 0.081 & $-$0.040 & $-$0.52 & 0.49 & $-$0.10 & 0.004 
  \end{tabular}
  \end{ruledtabular}
 \end{center}
\end{table}

The Hamiltonians for the single neutron $\hat H_\mathrm{F}^\nu$ 
and the boson-fermion interaction $\hat H_\mathrm{BF}^\nu$
are determined by using the method developed in 
Ref.~\cite{nomura2016odd}. 
The spherical single-particle energies 
$\epsilon_j$ and occupation probabilities $v_j^2$ of the odd-neutron 
orbital $j$ are provided by the same constrained SCMF calculations. 
In the following, since we consider for the fermionic 
degree of freedom only an odd neutron, the terms 
$\hat H_\mathrm{F}^\pi$ and $\hat H_\mathrm{BF}^\pi$ 
in Eq.~(\ref{eq:ham}), as well as the 
subscript $\rho$ in $j_\rho$'s are omitted. 
The strength parameters for the boson-fermion 
interaction $\hat H_{\mathrm{BF}}$, denoted by $\Gamma^{sdg}$, 
$\Lambda^{sdg}$, and $A^{sdg}$ ($\Gamma^{h}$, 
$\Lambda^{h}$, and $A^{h}$) for positive (negative) parity, 
are treated as the only free parameters, and 
are determined, separately for each parity, to reproduce the experimental 
low-lying excitation spectra. 
The criteria for fitting these parameters are that 
the spin of the ground state (i.e., the lowest-energy state for each parity) 
should be reproduced, as well as the excitation energies of few lowest yrast states 
to a reasonable accuracy. Of course, the overall systematics of the lowest bands, 
i.e., the energy level spacing within the bands and the observed 
$\Delta I=1$ or 2 systematics, should also  
be reproduced. Transition strengths are not taken into account in the fitting procedure. 

The adopted $\epsilon_j$ and $v^2_j$ for each orbital, 
and the boson-fermion interaction 
strengths are shown in Table~\ref{tab:vv} and 
Table~\ref{tab:ibfm2para}, respectively. 
As the strength parameters are adjusted for each 
odd-A nucleus, they should reflect the corresponding 
difference in structure between neighbouring isotopes. 
For instance, there are significant differences in 
these parameters between $^{95}$Zr and $^{97}$Zr 
both for the $sdg$ (positive-parity) and $h_{11/2}$ (negative-parity) 
configurations. 
%This indicates that $^{95}$Zr is near spherical and is 
%quite different in structure from the other odd-A Zr nuclei. 
In addition, one may notice in Table~\ref{tab:ibfm2para} 
that unusually large values of the exchange 
interaction strengths are chosen 
for the $1h_{11/2}$ configuration in $^{97-103}$Zr. 
In many IBFM calculations 
the typical value of this parameter is a few MeVs. In the 
present case the large values arise because the occupation 
probabilities for the 
$1h_{11/2}$ orbital obtained from the SCMF calculation are 
very small, e.g., $v^2_{h_{11/2}}=0.020$ for $^{97}$Zr 
(see Table~\ref{tab:vv}), 
and consequently the factor $\beta^2_{jj}\propto u_j^2v_j^2$ 
in Eq.~(\ref{eq:exc}) is also small. 
In order to account for the small $v^2_j$ values, 
a large strength for the exchange term 
$\Lambda$ is required specifically for the 
$1h_{11/2}$ configuration. 
In fact, the resulting constant 
$\Lambda_{jj}\equiv\beta_{jj}^2\sqrt{10/N_{\nu}(2j+1)}$ 
takes a realistic value, e.g., for $^{97}$Zr, for which 
the largest $\Lambda^h$ is obtained, it is approximately 
$\Lambda_{jj}=-2.2$ MeV. 
We also note that such large exchange strength parameters 
of the order $\Lambda\approx 50$ MeV were already considered 
in some previous studies, e.g., in Ref.~\cite{yoshida1994}. 

The resulting IBFM-2 Hamiltonian, with the parameters thus determined, 
is diagonalized to produce 
excitation energies and transition rates for a given odd-A nucleus.

%-----------------------------------------------------------------------
%
%	TABLE OF V^2 AND S.P.E.
%
%-----------------------------------------------------------------------
\begin{table}[htb!]
 \begin{center}
\caption{\label{tab:vv} 
Neutron single-particle energies $\epsilon_{j}$ (in MeV )
 and occupation probabilities $v^2_{j}$ obtained from spherical SCMF 
 calculations for the odd-A nuclei $^{95,97,99,101,103}$Zr. }
 \begin{ruledtabular}
  \begin{tabular}{ccccccc}
   & & $3s_{1/2}$ & $2d_{3/2}$ & $2d_{5/2}$ & $1g_{7/2}$ & $1h_{11/2}$ \\
\hline
$^{95}$Zr & $\epsilon_j$  & $-$4.322 & $-$3.838 & $-$6.219 & $-$5.199 & $-$0.894 \\
                 & $v^2_j$ & 0.078 & 0.068 & 0.484 & 0.204 & 0.014 \\
$^{97}$Zr & $\epsilon_j$  & $-$4.557 & $-$4.038 & $-$6.418 & $-$5.499 & $-$1.149 \\
                 & $v^2_j$ & 0.127 & 0.099 & 0.604 & 0.327 & 0.020 \\
$^{99}$Zr & $\epsilon_j$  & $-$4.773 & $-$4.241 & $-$6.614 & $-$5.801 & $-$1.409 \\
                 & $v^2_j$ & 0.188 & 0.135 & 0.699 & 0.462 & 0.026 \\
$^{101}$Zr & $\epsilon_j$  & $-$4.970 & $-$4.445 & $-$6.806 & $-$6.099 & $-$1.671 \\
                  & $v^2_j$ & 0.265 & 0.181 & 0.777 & 0.600 & 0.031 \\
$^{103}$Zr & $\epsilon_j$  & $-$5.146 & $-$4.643 & $-$6.993 & $-$6.388 & $-$1.929 \\
                  & $v^2_j$ & 0.367 & 0.245 & 0.843 & 0.730 & 0.036 \\
  \end{tabular}
   \end{ruledtabular}
 \end{center}
\end{table}

%-----------------------------------------------------------------------
%
%	TABLE OF BOSON-FERMION STRENGTHS
%
%-----------------------------------------------------------------------
\begin{table}[htb!]
 \begin{center}
\caption{\label{tab:ibfm2para} 
The adopted values for the boson-fermion strength parameters of 
the IBFM-2 Hamiltonian $\hat H_\mathrm{BF}$, used for the $sdg$ and 
$h_{11/2}$ configurations to describe the 
positive- and negative-parity low-lying states, respectively, of the odd-A nuclei 
$^{95-103}$Zr. All entries in the table are in the units of MeV.}
\begin{ruledtabular}
  \begin{tabular}{ccccccc}
   & $\Gamma^{sdg}$ & $\Lambda^{sdg}$ & $A^{sdg}$ 
   & $\Gamma^{h}$ & $\Lambda^{h}$ & $A^{h}$\\
\hline
$^{95}$Zr & 0.1 & 0.0 & 0.0 & 0.4 & 0.4 & 0.0 \\
$^{97}$Zr & 0.3 & 3.6 & 0.0 &  0.5 & 47.0 & 4.0 \\
$^{99}$Zr & 0.5 & 1.3 & $-$4.0 & 0.5 & 30.0 & $-$3.0 \\
$^{101}$Zr & 0.5 & 0.66 & $-$0.3 & 0.1 & 17.0 & $-$0.0 \\
$^{103}$Zr & 0.2 & 0.66 & 0.0 & 0.2 & 12.6 & 0.0
  \end{tabular}
  \end{ruledtabular}
 \end{center}
\end{table}

%-----------------------------------------------------------------------
%
%	EXCITATION SPECTRA FOR EVEN-EVEN NUCLEI
%
%-----------------------------------------------------------------------
\begin{figure}[htb!]
\begin{center}
\includegraphics[width=\linewidth]{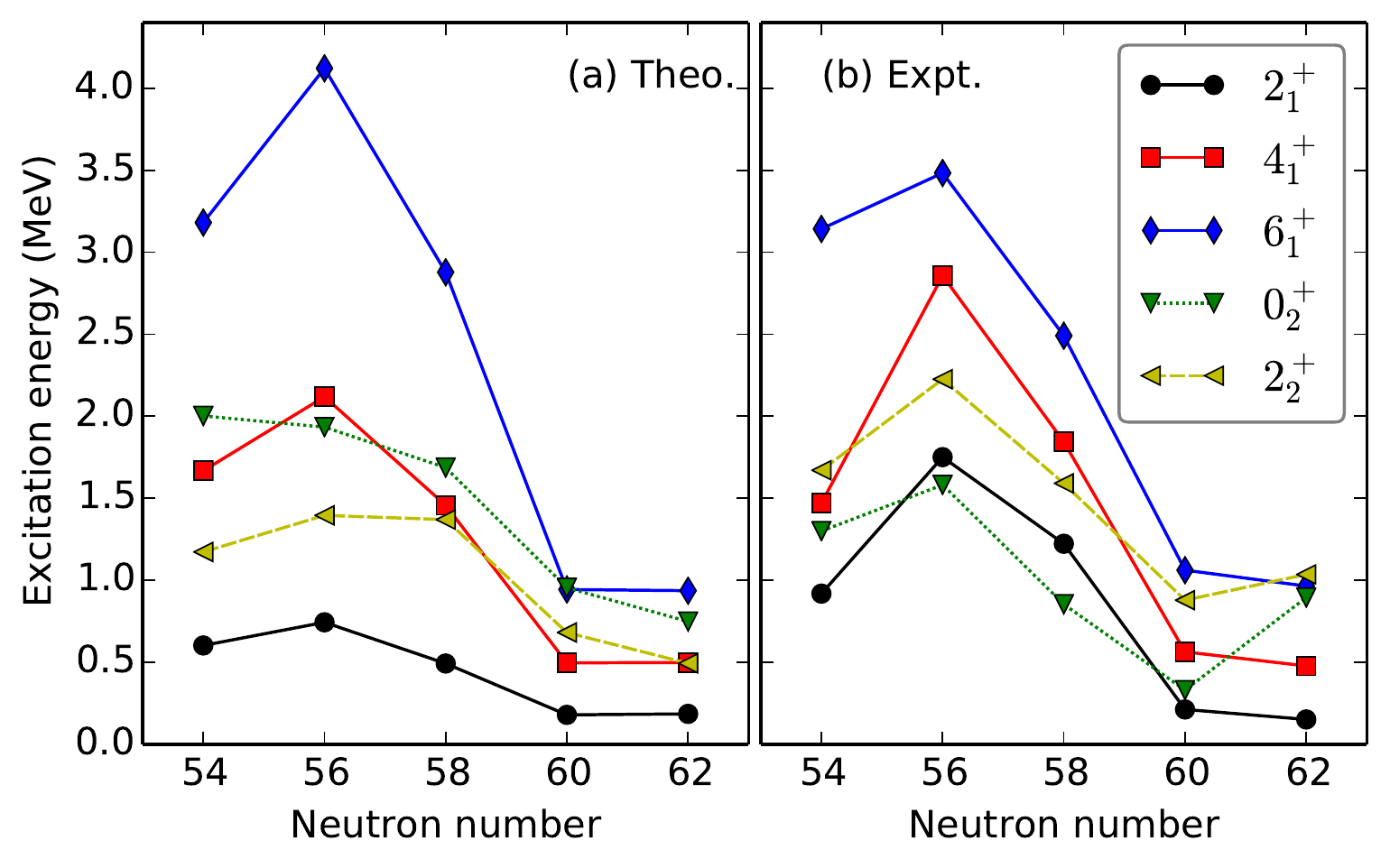}
\caption{(Color online) 
Low-energy excitation spectra of the 
even-even isotopes $^{94-102}$Zr, calculated with the IBM-2 Hamiltonian of Eq.~(\ref{eq:ibm2}). 
The corresponding data, taken from the compilation of the 
ENSDF database \cite{data}, are included for comparison.} 
\label{fig:level-even}
\end{center}
\end{figure}

%-----------------------------------------------------------------------
%
%	SPECTRA FOR ODD-A ZR NUCLEI
%
%-----------------------------------------------------------------------
\begin{figure*}[htb!]
\begin{center}
\includegraphics[width=0.8\linewidth]{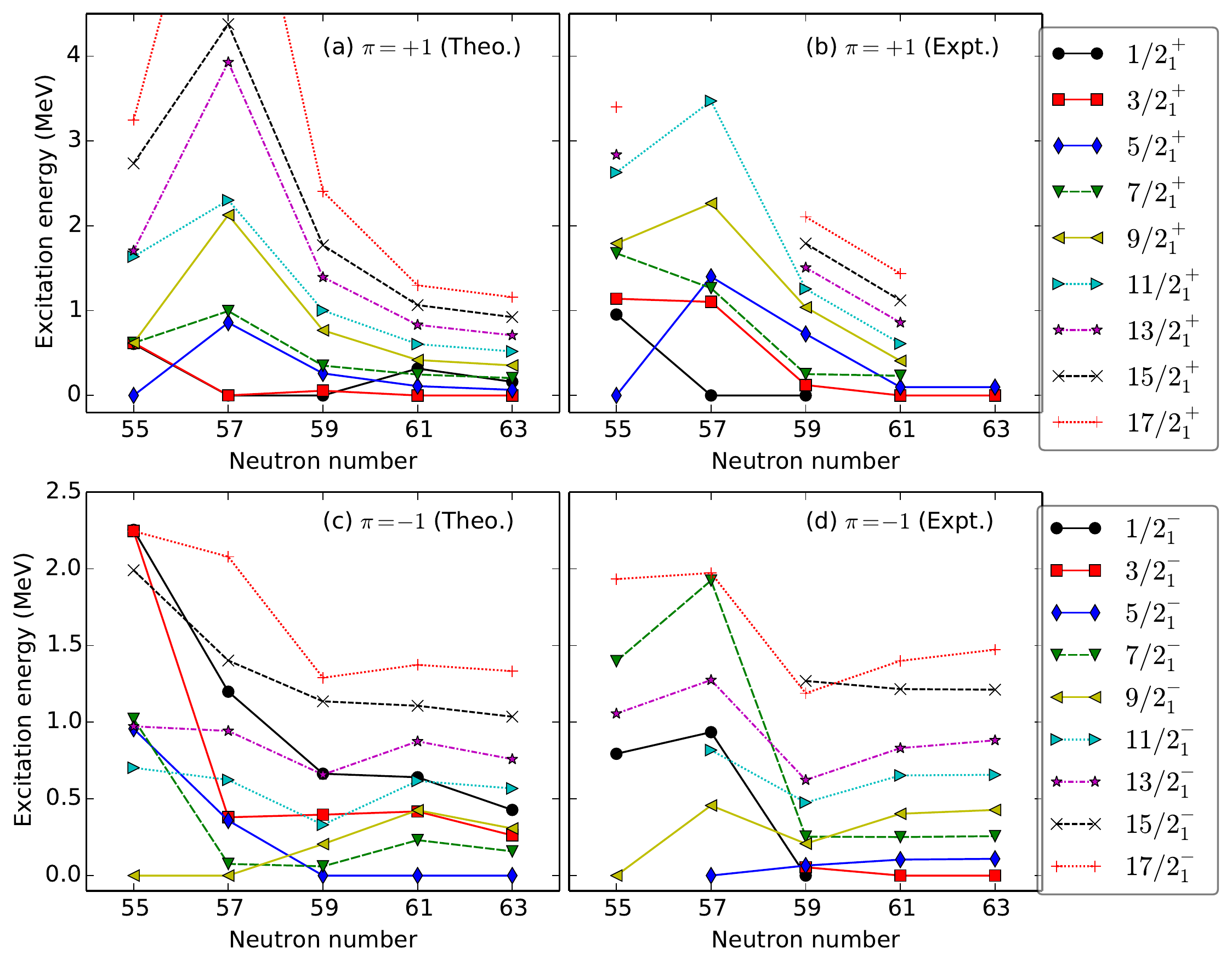} 
\caption{(Color online) Excitation spectra for the 
low-lying positive (a,b) and negative-parity (c,d) states 
of the odd-A nuclei $^{95-103}$Zr. The experimental levels are 
from Refs.~\cite{data,matejska-minda2009,rzacaurban2018,spagnoletti2019}.} 
\label{fig:level-odd}
\end{center}
\end{figure*}

\section{Spectroscopic properties\label{sec:results}}

\subsection{Excitation spectra of even-even Zr isotopes}

The energy spectra of low-lying excited states in the 
even-even Zr isotopes are depicted in Fig.~\ref{fig:level-even}. 
The transition between different shapes with increasing neutron number 
is characterized by the rapid decrease of the low-spin levels  
starting from $^{96}$Zr to $^{100}$Zr. 
The fact that the lowest levels for $^{96}$Zr are found 
at rather high energy when compared to the neighboring isotopes 
points to the $N=56$ neutron 
sub-shell closure (due to the filling of the $2d_{5/2}$ orbital), 
and the sudden decreases of the energy levels from 
$N=56$ towards $N=60$ corresponds to the enhancement of collectivity. 
As shown in Fig.~\ref{fig:level-even}, these empirical features are 
qualitatively reproduced by the present calculation.

One notices, however, that the 
excitation energies of the second $0^+$ state in the nuclei
$^{98,100}$Zr are predicted 
far too high with respect to their experimental counterparts. 
The occurrence of very low-lying excited $0^+$ states is often attributed to  
effects such as shape coexistence related to intruder configurations, and 
to pairing vibrations, both of which 
are outside the model space of the present IBM framework. 
Several recent IBM calculations 
\cite{nomura2016zr,gavrielov2019,garciaramos2019} 
that include the effects of intruder excitations 
across the proton $Z=40$ sub-shell closure 
and configuration mixing of normal and intruder configurations,  
reproduced the $0_2^+$ excitation energies. 
A drawback of such extended calculations is 
that, since two independent Hamiltonians associated 
with different boson numbers need to be introduced \cite{duval1981}, 
the number of model parameters increases significantly.  
In particular, the extension of this formalism to odd-mass systems, i.e., 
to the case of an odd nucleon coupled to the configuration-mixing IBM core, 
becomes exceptionally complex. 
The current implementation of the IBFM does not  
perform configuration mixing in the boson space and, therefore, here 
the calculation for the even-even Zr isotopes is carried out without 
the inclusion of intruder excitations and configuration mixing.

For $^{94,96}$Zr the present 
calculation predicts a 
level structure characterized by the energy ratio 
$E(4^+_1)/E(2^+_1)>2$. 
This is at variance with the experimental results, which exhibit 
a smaller ratio $E(4^+_1)/E(2^+_1)<2$. 
The discrepancy could be accounted for by the 
fact that the employed IBM consists of only collective 
nucleon pairs of monopole and quadrupole types 
(i.e., $s$ and $d$ bosons). 
For the transitional nucleus $^{98}$Zr, the $2^+_1$ 
level is particularly low, as in the case of $^{96}$Zr. 
In our calculation the lowest-lying states for $^{98}$Zr 
are mostly based on configurations located close to the 
prolate global minimum at $\beta\approx 0.5$ 
on the SCMF energy surface. 
The resulting IBM spectra are likely to be more 
rotational-like than observed in experiment. 
We obtain $\gamma$-soft spectra 
for $^{100,102}$Zr, and this result is consistent 
with the underlying SCMF surfaces, which are indeed 
soft in the $\gamma$ degree of freedom.

\subsection{Excitation spectra of odd-A Zr isotopes}

The principal scope of this work are spectroscopic 
calculations of structural evolution in the odd-A Zr isotopes, 
and in the following we discuss in much more detail the results for 
odd-A systems. 
Firstly, in Fig.~\ref{fig:level-odd} we display the systematics of 
calculated excitation spectra for the low-lying positive- 
and negative-parity yrast states of the odd-A Zr isotopes, 
in comparison to available data 
\cite{data,matejska-minda2009,rzacaurban2018,spagnoletti2019}. 
The excitation energies of negative-parity states are plotted 
with respect to the energy of the lowest-lying negative parity state. 
One notices that the calculated spectra 
reproduce very nicely the experimental results 
for both parities, except perhaps for 
the excitation energy of ${3/2}^-$ in most of the odd-A Zr.

For both parities the level structure 
changes significantly between $^{97}$Zr and $^{101}$Zr. 
The fact that the experimental spectra are particularly expanded at $N=57$, 
that is, the excitation energies of most levels exhibit peaks at $N=57$, is interpreted 
as an effect of the neutron $2d_{5/2}$ sub-shell filling in the corresponding 
even-even core nucleus $^{96}$Zr. 
The calculated positive-parity states are in better agreement with experiment compared 
to the negative-parity states, in particular at $N=57$. This is probably because   
for the negative parity only the unique-parity $1h_{11/2}$ orbital is considered. 
For the lighter odd-A Zr isotopes the 
energy spectra of $^{95,97}$Zr appear almost harmonic. 
In the transitional region at $^{99}$Zr many of the yrast levels 
are lowered in energy, and a more complicated 
low-lying structure with higher level density emerges. 
For negative-parity states, in particular, many of the higher-spin levels 
exhibit a sharp lowering in energy at the neutron number $N=59$. 
For the heavier isotopes $^{101,103}$Zr, we find a more regular 
pattern of excitation spectra, characterized by the $\Delta I=1$ 
level sequence with increasing angular momentum. 
As one notices from Fig.~\ref{fig:level-odd} (c,d), in most odd-A nuclei
the spin of the calculated lowest negative-parity state is at variance with 
data. This could be due to the calculated occupation number $v^2_{h_{11/2}}$ and the 
resulting boson-core interaction. However, we also note that 
the lowest-state spins for the negative-parity states are, in many cases, 
not firmly established experimentally \cite{data}.

%-----------------------------------------------------------------------
%
%	LEVEL SCHEME: 97ZR
%
%-----------------------------------------------------------------------
\begin{figure}[htb!]
\begin{center}
\includegraphics[width=\linewidth]{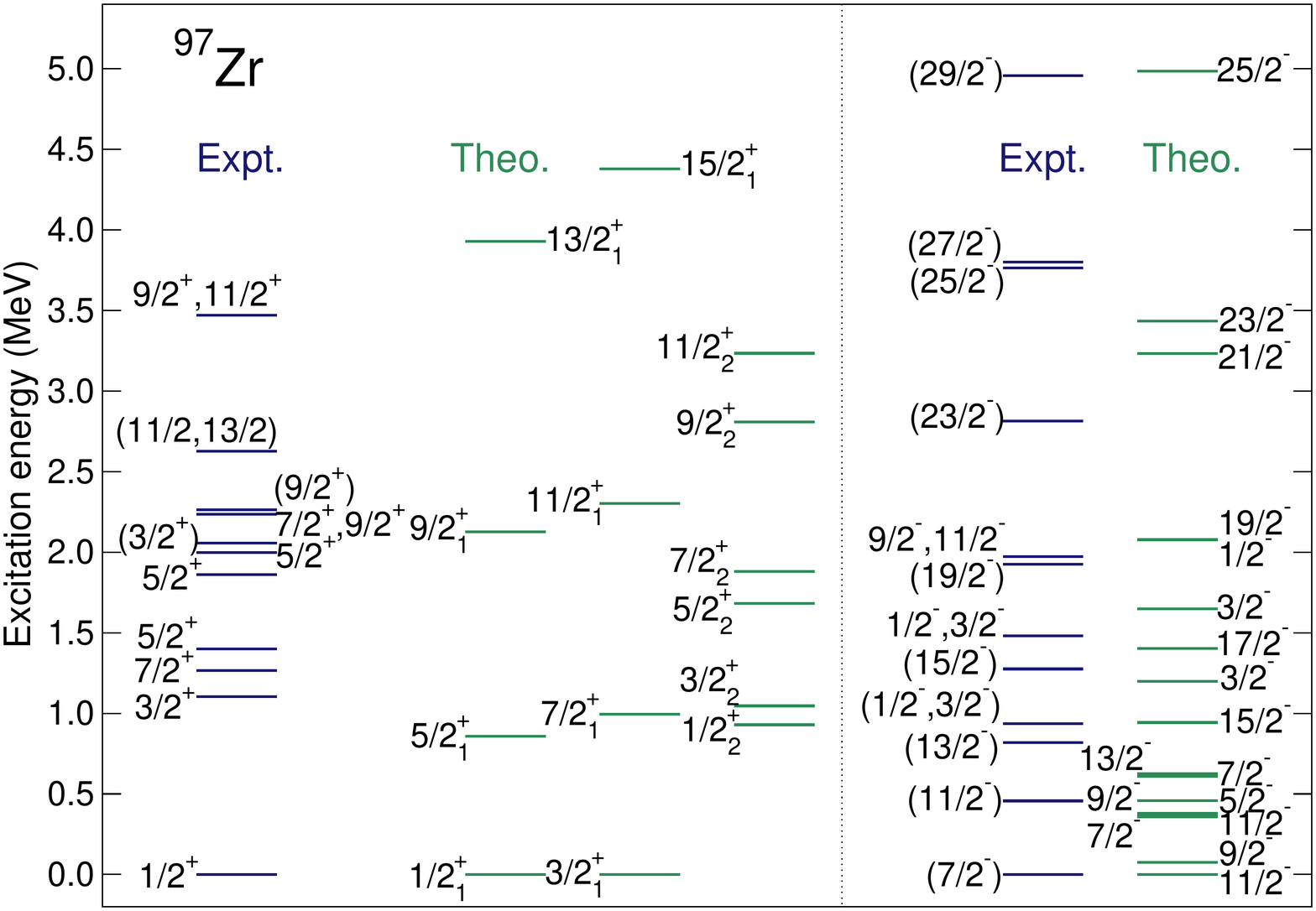} 
\caption{(Color online) Comparison between theory and experiment \cite{rzacaurban2018,data} for the 
positive (left) and negative-parity (right) excitation spectrum of $^{97}$Zr.} 
\label{fig:zr97}
\end{center}
\end{figure}

%-----------------------------------------------------------------------
%
%	LEVEL SCHEME: 99ZR
%
%-----------------------------------------------------------------------
\begin{figure*}[htb!]
\begin{center}
\includegraphics[width=0.6\linewidth]{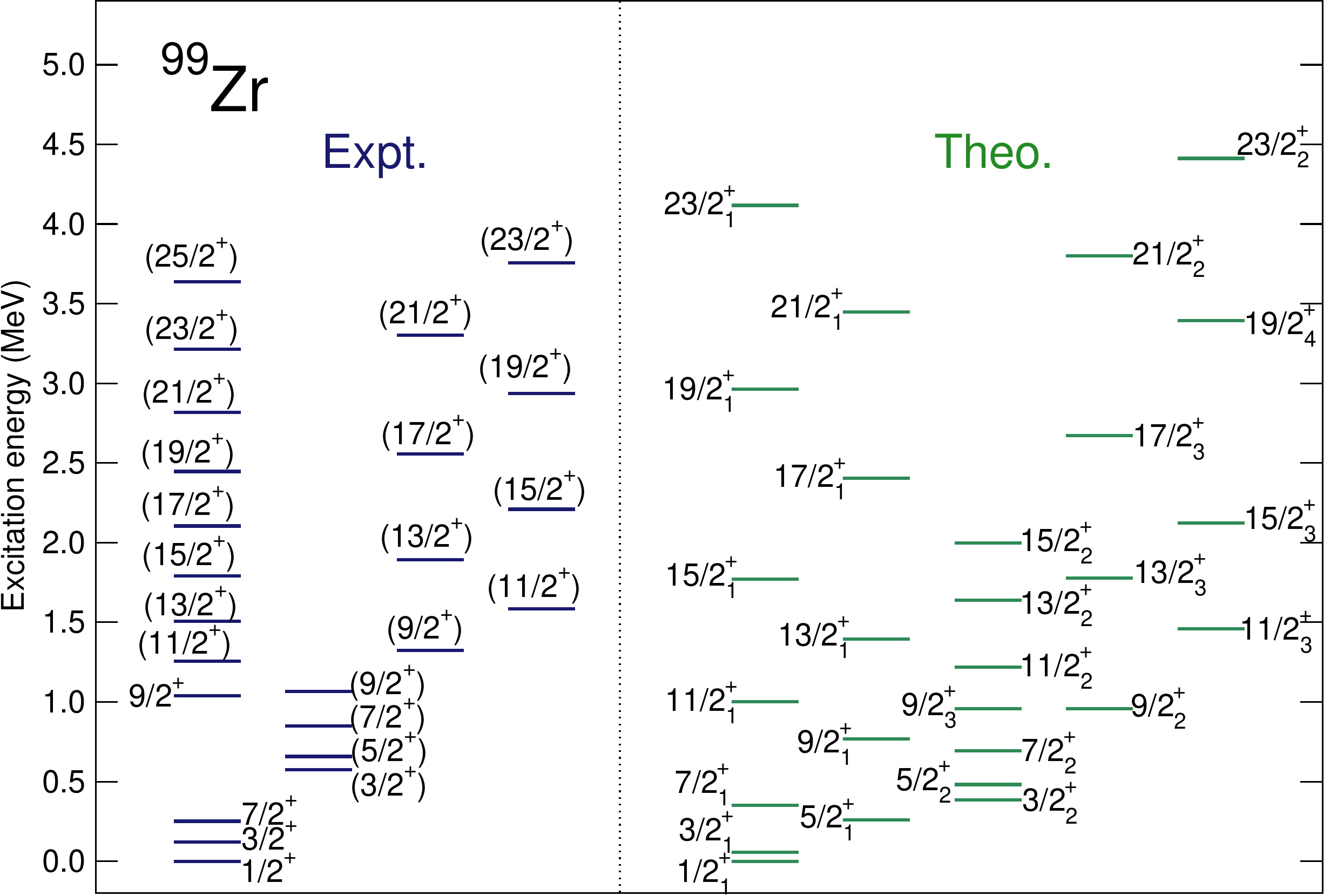} \\
\includegraphics[width=0.6\linewidth]{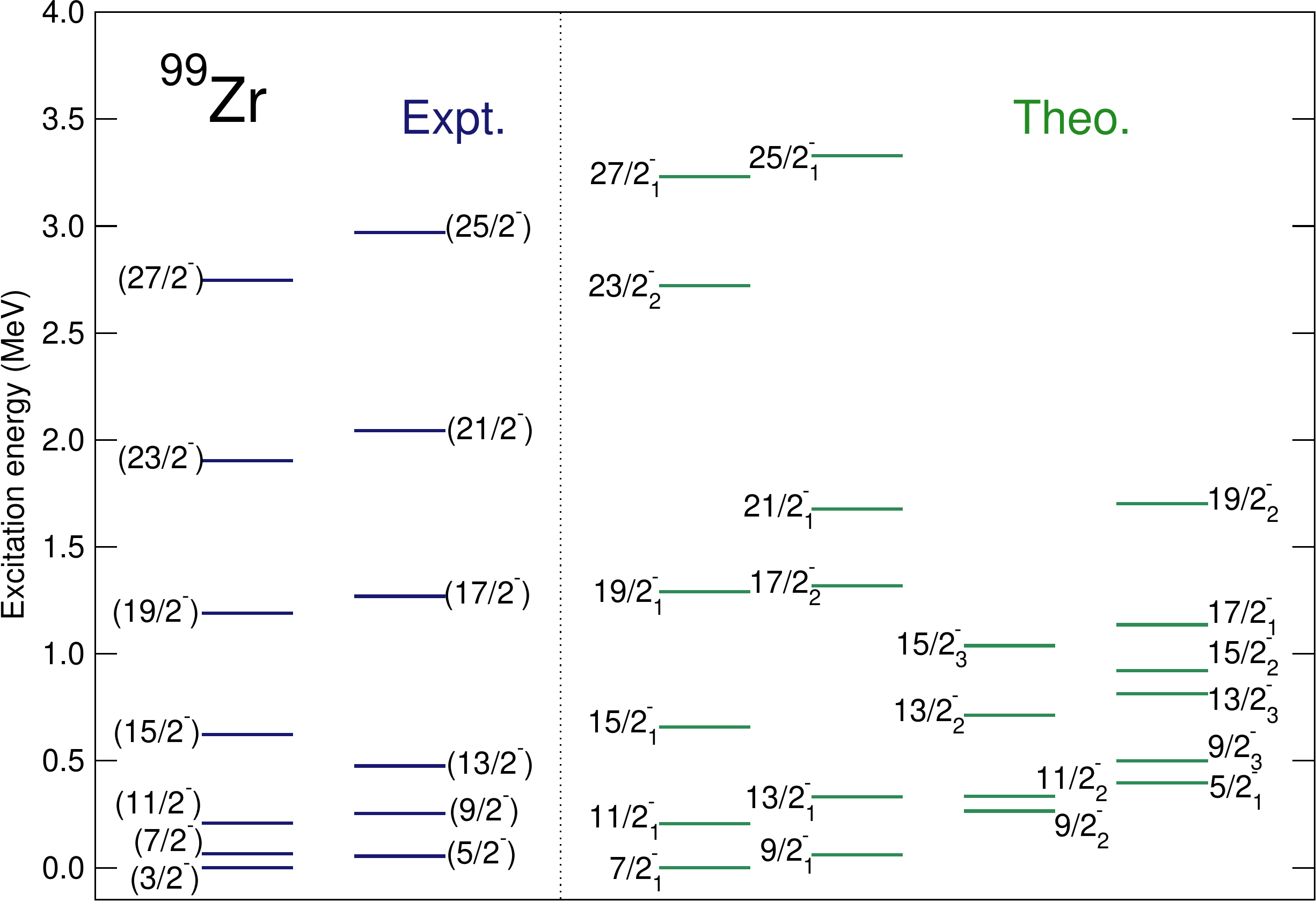}
\caption{(Color online) Band structure of 
theoretical positive (upper panel) and negative-parity (lower panel) excitation spectra of 
$^{99}$Zr in comparison to the available data \cite{spagnoletti2019,data}.} 
\label{fig:zr99}
\end{center}
\end{figure*}

%-----------------------------------------------------------------------
%
%	LEVEL SCHEME: 101ZR
%
%-----------------------------------------------------------------------
\begin{figure*}[htb!]
\begin{center}
\includegraphics[width=0.6\linewidth]{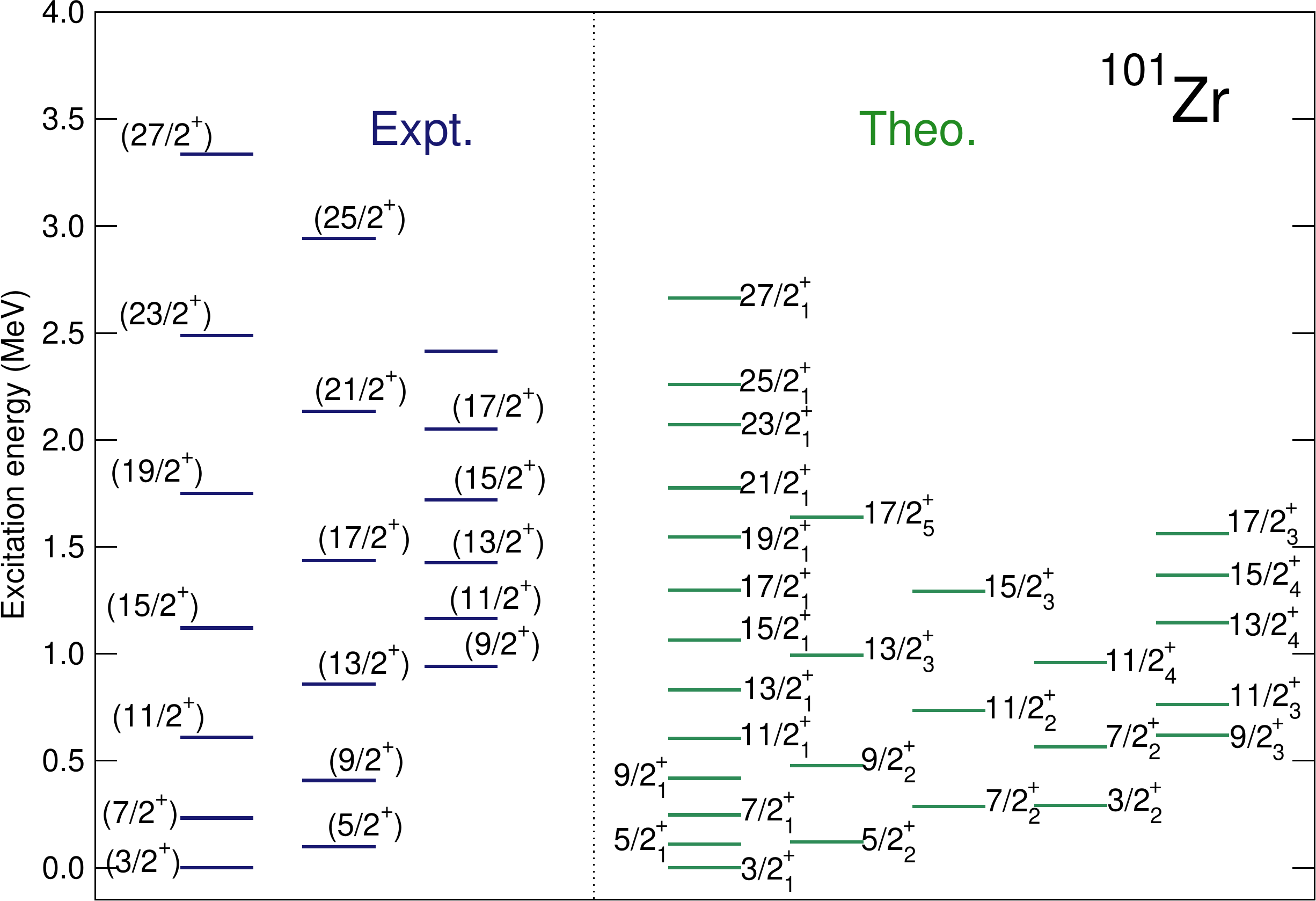} \\ 
\includegraphics[width=0.6\linewidth]{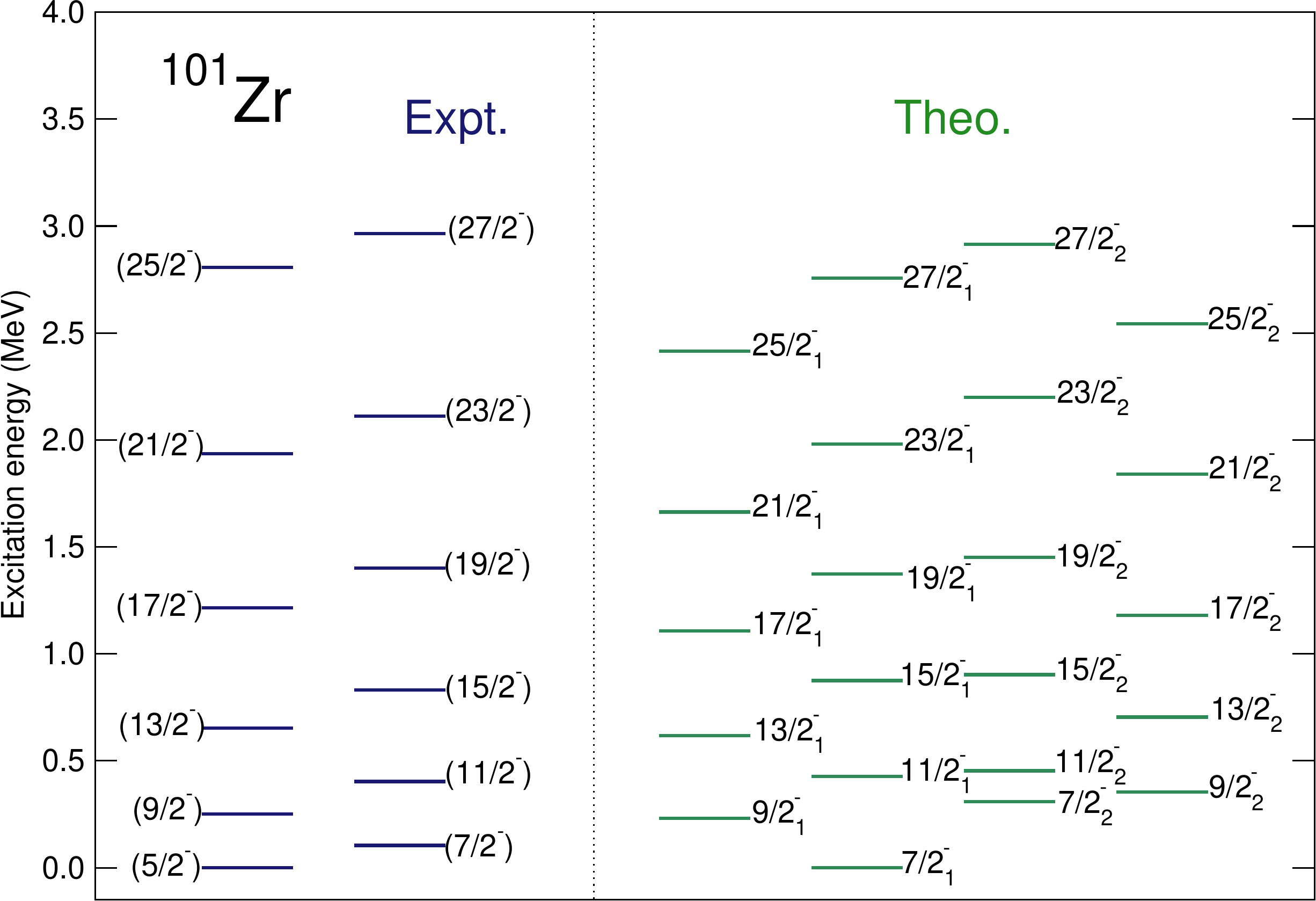}
\caption{(Color online) Same as in the caption to Fig.~\ref{fig:zr99}, but for the nucleus $^{101}$Zr. 
The data are from Ref.~\cite{data,orlandi2006}.} 
\label{fig:zr101}
\end{center}
\end{figure*}

%-----------------------------------------------------------------------
%
%	WAVE FUNCTION CONTENTS
%
%-----------------------------------------------------------------------
\begin{figure}[htb!]
\begin{center}
\includegraphics[width=\linewidth]{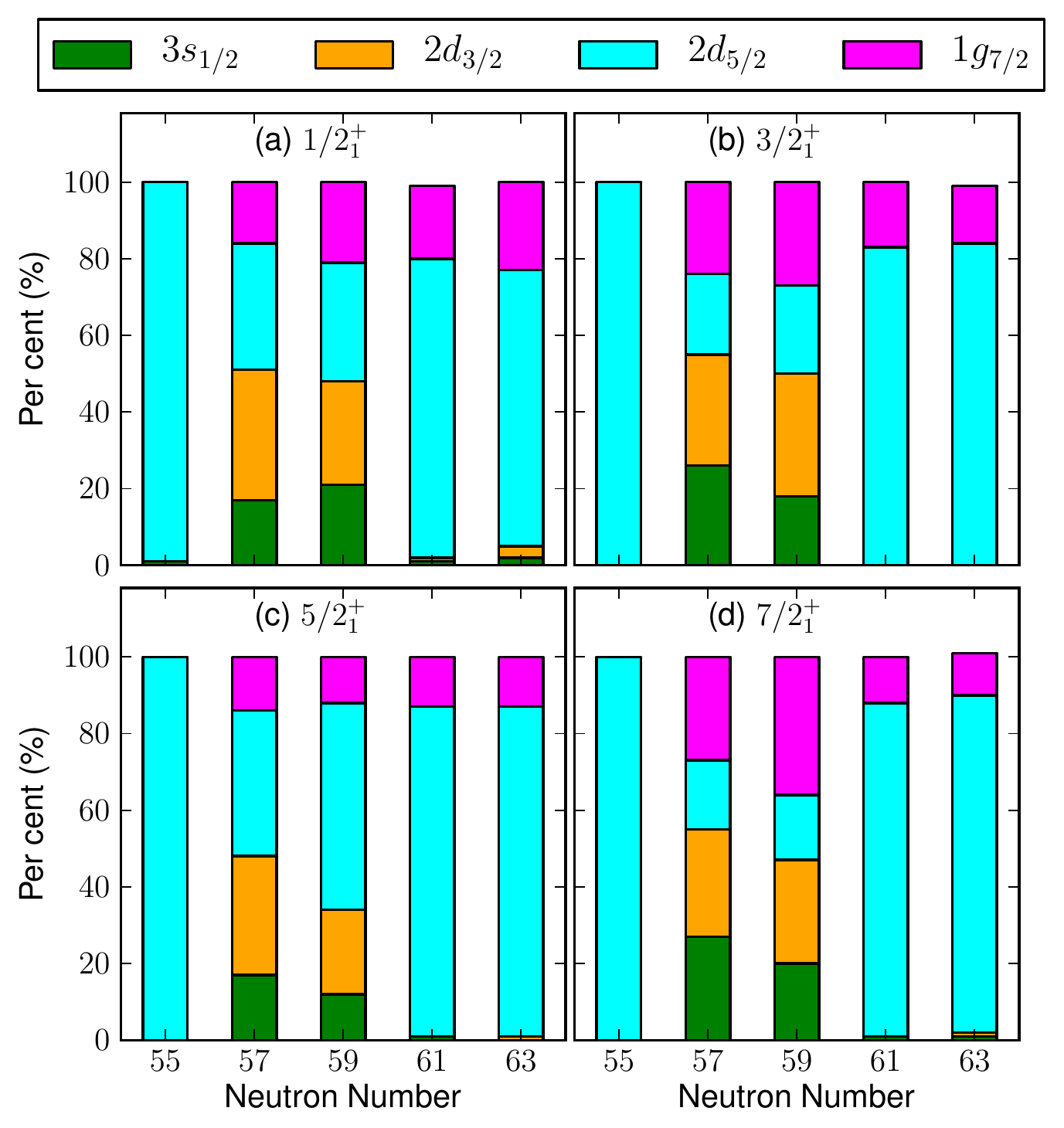} 
\caption{(Color online) Probability amplitudes 
of the $3s_{1/2}$, $2d_{3/2}$, $2d_{5/2}$, and $1g_{7/2}$ 
single-neutron configurations in the wave functions of the 
calculated positive-parity yrast states ${1/2}^+_1$ (a), 
${3/2}^+_1$ (b), ${5/2}^+_1$ (c), and ${7/2}^+_1$ (d) 
in the odd-A isotopes $^{95-103}$Zr.} 
\label{fig:wf}
\end{center}
\end{figure}

\subsection{Detailed level schemes of selected odd-A Zr nuclei}

It is interesting to consider in more detail the excitation spectra of individual odd-A 
Zr isotopes in the transitional region. 
Figures~\ref{fig:zr97}, \ref{fig:zr99}, and \ref{fig:zr101} display the 
lowest band structures of both parities in $^{97,99,101}$Zr, 
which are most relevant for the discussion of a shape transition. 
Included are also the corresponding experimental spectra for comparison. 
To help with the analysis of the structure of the lowest positive-parity states, 
in Fig.~\ref{fig:wf} we plot the the probability amplitudes of the 
$3s_{1/2}$, $2d_{3/2}$, $2d_{5/2}$, and $1g_{7/2}$ 
single-particle configurations in the wave functions 
of the yrast states ${1/2}^+_1$, ${3/2}^+_1$, ${5/2}^+_1$, and ${7/2}^+_1$. 

\subsubsection{$^{97}$Zr}

There is no definite band structure established experimentally 
in $^{97}$Zr. As it can be deduced from Fig.~\ref{fig:wf}, 
it appears that all four single-particle configurations 
($3s_{1/2}$, $2d_{3/2}$, $2d_{5/2}$, and $1g_{7/2}$) 
almost equally contribute to the composition of the wave functions of the 
lowest-lying positive-parity states. 
Our calculation predicts two $\Delta I=2$ positive-parity bands characteristic 
for the weak-coupling limit, and a $\Delta I=1$ band with a pronounced doublet 
structure built on the ${1/2}^+_2$. 
At variance with the data, the ${3/2}^+_1$ state is calculated 
too low in energy, just a few keV above the ${1/2}^+$ ground state. 
For the negative-parity states, the calculation 
predicts many more levels than observed in experiment so far, and 
also the E2 strengths of these states are strongly fragmented. 
This makes the assignment of low-lying negative-parity 
states into bands almost impossible.

\subsubsection{$^{99}$Zr}

Unlike $^{97}$Zr, several band structures have recently been experimentally identified  
in the nucleus $^{99}$Zr \cite{spagnoletti2019,boulay2020,data}.  
Both the experimental and theoretical positive-parity 
energy spectra in Fig.~\ref{fig:zr99} exhibit strongly-coupled 
$\Delta I=1$ and weakly-coupled $\Delta I=2$ bands coexisting 
at low energy. 
As seen in Fig.~\ref{fig:wf}, the structure of the low-lying 
low-spin positive-parity yrast states is similar 
to that of $^{97}$Zr: all four single-particle configurations equally contribute  
to the IBFM-2 wave functions. For instance, in the ${1/2}^+_1$ ground state 
the $3s_{1/2}$, $2d_{3/2}$, $2d_{5/2}$, and $1g_{7/2}$ single-particle 
configurations contribute with probabilities of 21 \%, 27 \%, 31 \%, and 21 \%, 
respectively. 
In contrast, most of the states in the  
$\Delta I=1$ band based on the ${3/2}^+_2$ state, are predominantly  
(about 80 \%) composed 
of the $2d_{5/2}$ single-particle configuration. 
Another two $\Delta I=2$ weakly-coupled 
bands built on top of the ${9/2}^+_2$ and ${11/2}^+_3$ 
states are predicted. The main component of these bands is, 
again, the $2d_{5/2}$ configuration, especially for higher-spin 
states in the bands. In the lower-spin states close the 
${9/2}^+_2$ and ${11/2}^+_3$ band-heads, the four single-particle 
configurations are so strongly mixed, that the band assignment 
for these states according to the systematics of the 
E2 transitions is not very certain. 

The experimental negative-parity spectra look 
much more regular, with only two $\Delta I=2$ bands extending to high-spin. 
The calculation reproduces the overall structure of the experimental negative-parity spectra, but does not confirm 
the assigned band-heads of the two $\Delta I=2$.

\subsubsection{$^{101}$Zr}

The even-even core for this nucleus ($^{100}$Zr) 
is located near the end of the phase transition, and 
the $(\beta,\gamma)$ energy surface exhibits a more  
extended prolate deformation at large $\beta$. 
In contrast to $^{97,99}$Zr, the lowest-lying positive-parity 
states for $^{101}$Zr are predominantly composed 
of the $1g_{7/2}$ ($\approx 20 \%$) 
and $2d_{5/2}$ ($\approx 80 \%$) single-particle 
configurations (see Fig.~\ref{fig:wf}). 
The excitation spectra for both parities display a more 
regular band structure compared to $^{97,99}$Zr, 
and the states in each band are connected by strong E2 transitions. 
The calculated yrast band built on the ${3/2}^+_1$ ground state follows 
the strong-coupling $\Delta I=1$ systematics of the 
E2 transitions. 
%The two theoretical lowest $\Delta I=2$ band are also 
%connected by the strong E2 transitions. 
The second excited band in experiment, based on 
the tentatively assigned ${9/2}^+$ state at 940 keV, 
could be compared with the predicted strong-coupling 
$\Delta I=1$ band built on the ${9/2}^+_3$ state at 619 keV. 
However, one should keep in mind that this band has been  
assigned to the state ${9/2}[404]$ associated 
with the proton $1g_{9/2}$ intruder state \cite{orlandi2006},  
whereas this state is not included in the configuration space of the 
present IBFM-2 calculation. 

For the negative-parity two $\Delta I=2$ structures have been empirically 
identified as yrast bands. Several $\Delta I=2$ 
bands are also obtained in the calculation. The lowest two reproduce the excitation energies 
of the experimental bands but differ in spin by one unit. 
As mentioned above, this can partly be due to the limited IBFM-2 space that includes only 
the $1h_{11/2}$ negative-parity orbital. 
Note, however, that the spin assignment for 
the experimental states is tentative.
Also the theoretical band assignment in this case may not be unique, since several 
states with the same spin are calculated within a small energy 
interval and, because of mixing, their E2 transitions are   
weak and fragmented. The band structure for the neighbouring nucleus $^{103}$Zr 
is similar to the one obtained for $^{101}$Zr, but is not discussed 
here since there are no data available.

\subsection{Electromagnetic properties}

There is also limited experimental information about the 
electromagnetic transition rates for the odd-A Zr isotopes. 
These properties are readily computed using the eigenstates 
of the IBFM-2 Hamiltonian. The E2 operator $\hat T^\mathrm{(E2)}$ 
in the IBFM-2 takes the form 
\cite{IBFM}:
\begin{align}
 \label{eq:e2}
\hat T^\mathrm{(E2)}
= e_\nu^\mathrm{B}\hat Q_\nu + e_\pi^\mathrm{B}\hat Q_\pi
-\frac{1}{\sqrt{5}}e^\mathrm{F}
\sum_{jj'}
\gamma_{jj'}
%(u_{j}u_{j'}-v_{j}v_{j'})
%\braket{ j' \| e^\mathrm{F}r^2Y^{(2)} \| j}
(a_{j}^\dagger\times\tilde a_{j'})^{(2)},
%\nonumber \\
\end{align}
where the fixed values for the boson effective charges 
$e^\mathrm{B}_\nu = e^\mathrm{B}_\pi =0.10$ $e$b 
are chosen so that 
the $B({\mathrm E2}; 2^+_1\rightarrow 0^+_1)$ values 
for the deformed even-even core nuclei, i.e., $^{100,102}$Zr, 
are reproduced. 
The neutron effective charge $e^\mathrm{F} =0.5$ $e$b 
is adopted from our earlier calculation \cite{nomura2020cs}. 
%so that the $B$(E2) rates between the low-spin states 
%in the odd-A Zr nuclei are in an overall good agreement with the 
%experimental data.  
The M1 transition operator $\hat T^\mathrm{(M1)}$ reads 
\begin{align}
 \label{eq:m1}
\hat T^\mathrm{(M1)}
&=\sqrt{\frac{3}{4\pi}}
\Biggl\{
g_\nu^\mathrm{B}\hat L^\mathrm{B}_\nu + g_\pi^\mathrm{B}\hat
L^\mathrm{F}_\pi
-\frac{1}{\sqrt{3}}
\sum_{jj'}
\nonumber \\
&\times(u_{j}u_{j'}+v_{j}v_{j'})
\braket{
j' \| g_l^\nu{\bf l}+g_s^\nu{\bf s} \| j}
(a_{j}^\dagger\times\tilde
a_{j'})^{(1)}
\Biggr\}. 
%\nonumber \\
\end{align}
The empirical $g$-factors for the neutron and
proton bosons, $g_\nu^\mathrm{B}=0\,\mu_N$ and 
$g_\pi^\mathrm{B}=1.0\,\mu_N$, respectively, are adopted. 
For the neutron $g$-factors, the standard Schmidt values 
$g_l^\nu=0\,\mu_N$ and $g_s^\nu=-3.82\,\mu_N$
%($g_l^\pi=1.0\,\mu_N$ and $g_s^\pi=5.58\,\mu_N$) 
are used, with $g_s$ quenched by 30\% with respect to the free value.

In Table~\ref{tab:em} we list the calculated $B$(E2) and $B$(M1) 
transition rates, the electric quadrupole $Q(I)$ and 
magnetic dipole $\mu(I)$ moments for the odd-A nuclei $^{95,97,99,101}$Zr, 
for which data are available. 
Only the quadrupole and magnetic moments for the 
ground state are known for $^{95}$Zr. The calculated $Q({5/2}^+_1)$ is 
rather small in magnitude. It is opposite in sign to the experimental value, 
which is, however, also relatively small in magnitude. 
The sign of the magnetic moment of $^{95}$Zr has not been identified experimentally, 
but it is likely to be negative from the present calculation. 
For the $^{97}$Zr, all the calculated experimental transition strengths 
and moments are in a good agreement with the data. 

The $B(E2; {7/2}^+_1\rightarrow {3/2}^+_1)$ transition rate in 
$^{99}$Zr is experimentally suggested to be rather weak \cite{boulay2020}, 
similar to the neighbouring isotope $^{97}$Zr. 
The predicted E2 strength for this transition is a bit larger, 
but is in the same order of magnitude as the experimental one. 
The experimental $B(E2; {7/2}^+_2\rightarrow {3/2}^+_2)$ transition 
rate of $46\pm 12$ W.u. is considerably underestimated by the 
calculation. 
As seen in Fig.~\ref{fig:zr99}, both the ${7/2}^+_2$ and ${3/2}^+_2$ 
states are in the same band in our calculation. This band 
is dominated by the $\Delta I=1$ E2 systematics, and the 
$\Delta I=2$ E2 transitions within the band are much weaker. 
The phenomenological IBFM calculation performed 
in Ref.~\cite{spagnoletti2019} has also underestimated 
the measured value of this transition strength 
by a factor of five. 
%$B(E2; {7/2}^+_2\rightarrow {3/2}^+_2)$ 
%transition
In the present calculation the $B$(E2) values for the 
negative-parity states in $^{99}$Zr are also by a 
factor of five to six lower than the experimental ones \cite{boulay2020}. 
Nevertheless, the majority of the $B$(M1) values, 
as well as the magnetic moments for the low-lying positive-parity 
states, both the sign and magnitude, are nicely reproduced. 

One notices that the electromagnetic properties 
for $^{101}$Zr are, overall, reasonably reproduced. 
The exceptions are perhaps the $B(E2; {7/2}^+_1\rightarrow {3/2}^+_1)$ 
rate, and few small magnetic moments that 
are obtained with the wrong sign. 

%-----------------------------------------------------------------------
%
%	E.M. TRANSITIONS TABLE
%
%-----------------------------------------------------------------------
\begin{table}[!htb]
\begin{center}
\caption{\label{tab:em} Calculated and experimental $B$(E2) and
 $B$(M1) transition rates (in Weisskopf units), and quadrupole $Q(I)$
 (in units of $e$b) and magnetic $\mu(I)$ (in units of $\mu_N$) moments for
 the odd-A nuclei $^{95,97,99,101}$Zr. The experimental values are 
 from Refs.~\cite{spagnoletti2019,boulay2020,data,stone2005}.}
 \begin{ruledtabular}
 \begin{tabular}{llrr}
 &  & Theory & Experiment \\
\hline
$^{95}$Zr 
& $Q({5/2}^+_1)$ & $-0.021$ & $+0.22(2)$ \\
& $\mu({5/2}^+_1)$ & $-1.33$ & 1.13(2) \\
$^{97}$Zr
& $B(E2; {5/2}^+_1\rightarrow {1/2}^+_1)$ & 6.2 & $>0.30$ \\
& $B(E2; {7/2}^+_1\rightarrow {3/2}^+_1)$ & 8.8 & 1.55(5) \\
& $B(E2; {11/2}^-_1\rightarrow {7/2}^-_1)$ & 0.15 & 0.25(6) \\
& $\mu({1/2}^+_1)$ & $-0.33$ & $-0.937(5)$ \\
& $\mu({7/2}^+_1)$ & $+2.54$ & $+1.37(14)$ \\
$^{99}$Zr
& $B(E2; {7/2}^+_1\rightarrow {3/2}^+_1)$ & 9.9 & 1.16(3) \\
& $B(E2; {7/2}^+_2\rightarrow {3/2}^+_2)$ & 2.9 & 46(12) \\
& $B(E2; {7/2}^-_1\rightarrow {3/2}^-_1)$ & 0.24 & $2.1\times 10^2$(7) \\
& $B(E2; {11/2}^-_1\rightarrow {7/2}^-_1)$ & 16 & 99(6) \\
& $B(E2; {15/2}^-_1\rightarrow {11/2}^-_1)$ & 12 & 60(11) \\
& $B(E2; {19/2}^-_1\rightarrow {15/2}^-_1)$ & 8.6 & 66(9) \\
& $B(M1; {3/2}^+_1\rightarrow {1/2}^+_1)$ & 0.0057 & 0.0102(3) \\
& $B(M1; {5/2}^+_1\rightarrow {3/2}^+_1)$ & 0.074 & 0.042(21) \\
& $B(M1; {5/2}^+_1\rightarrow {3/2}^+_2)$ & 0.0040 & 0.0047(20) \\
& $B(M1; {7/2}^+_2\rightarrow {5/2}^+_1)$ & 0.0098 & 0.032(10) \\
& $B(M1; {5/2}^-_1\rightarrow {3/2}^-_1)$ & 0.0063 & 0.015(9) \\
& $\mu({1/2}^+_1)$ & $-0.48$ & $-0.930(4)$ \\
& $\mu({3/2}^+_1)$ & $+0.75$ & $+0.42(6)$ \\
& $\mu({7/2}^+_1)$ & $+1.21$ & $\pm 2.31(14)$ \\
$^{101}$Zr
& $B(E2; {5/2}^+_1\rightarrow {3/2}^+_1)$ & 69 & $3.E+2^{4}_{3}$ \\
& $B(E2; {7/2}^+_1\rightarrow {3/2}^+_1)$ & 0.00061 & $>1.3\times 10^2$ \\
& $B(E2; {7/2}^-_1\rightarrow {5/2}^-_1)$ & 102 & $4.E+2(5)$ \\
& $B(M1; {5/2}^+_1\rightarrow {3/2}^+_1)$ & 0.017 & 0.036(13) \\
& $B(M1; {7/2}^+_1\rightarrow {5/2}^+_1)$ & 0.16 & $>0.091$ \\
& $B(M1; {7/2}^-_1\rightarrow {5/2}^-_1)$ & 0.15 & 0.033 \\
& $Q({3/2}^+_1)$ & $+0.70$ & $+0.81(6)$ \\
& $\mu({3/2}^+_1)$ & $-0.09$ & $-0.272(1)$ \\
& $\mu({5/2}^+_1)$ & $-0.51$ & $+0.117(65)$ \\
& $\mu({7/2}^+_1)$ & $-0.13$ & $<+0.59(50)$\\
& $\mu({5/2}^-_1)$ & $-1.33$ & $-0.50(23)$ \\
& $\mu({7/2}^-_1)$ & $-1.15$ & $-0.14(11)$ 
%\footnotetext[1]{Extracted from the measured $g$-factors \cite{orlandi2006}.}
%\footnotetext[2]{Extracted from the measured $g$-factors \cite{data}.}
 \end{tabular}
 \end{ruledtabular}
\end{center} 
\end{table}

%-----------------------------------------------------------------------
%
%	Q-INVARIANTS EVEN-EVEN ZR
%
%-----------------------------------------------------------------------
\begin{figure}[htb!]
\begin{center}
\includegraphics[width=\linewidth]{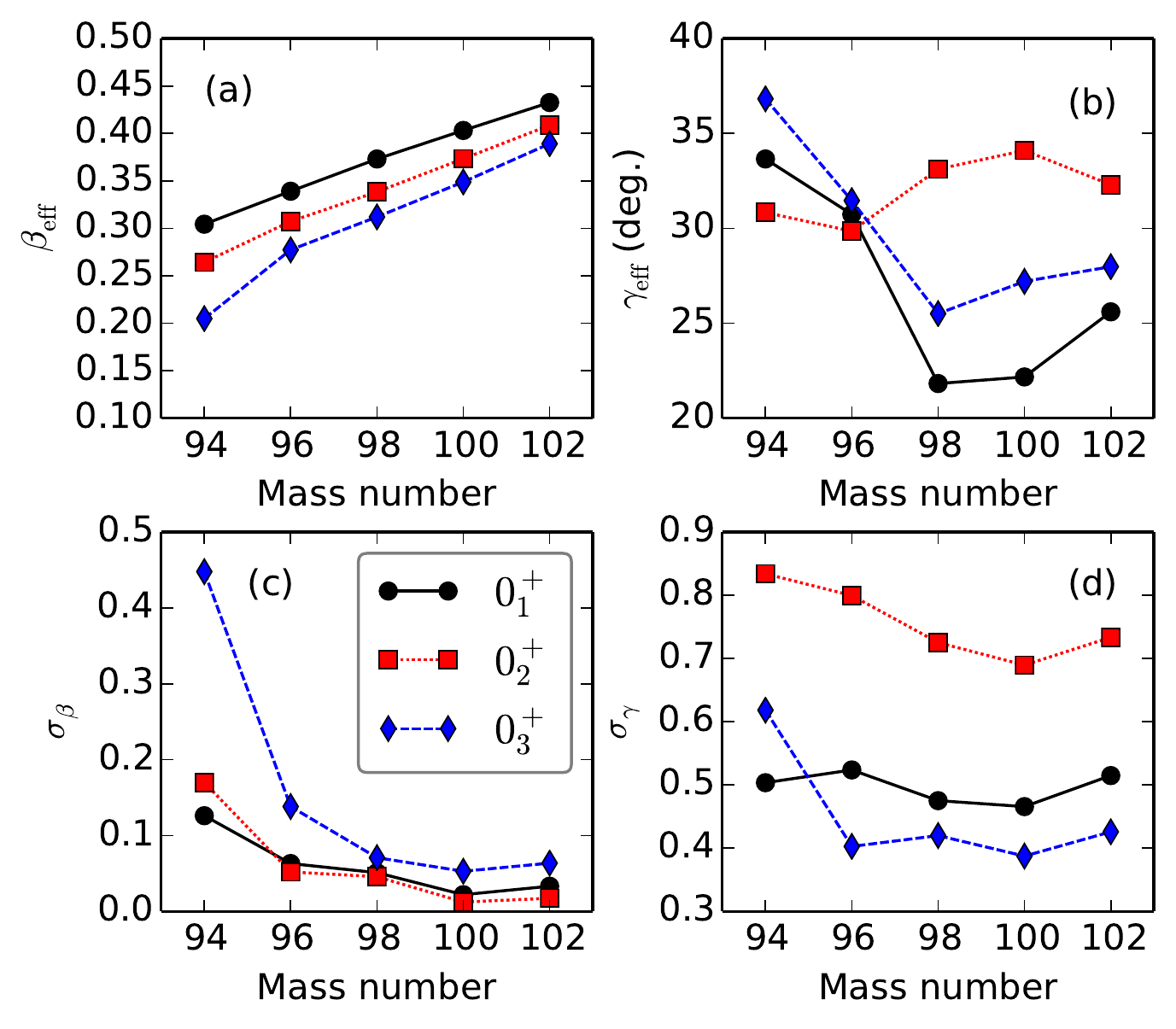} 
\caption{(Color online) The effective quadrupole deformation parameters 
$\beta_\mathrm{eff}$ (a) and $\gamma_\mathrm{eff}$ (b), 
and the fluctuations $\sigma_{\beta}$ (c) and $\sigma_{\gamma}$ (d), calculated 
for the lowest three $0^+$ states of the even-even Zr nuclei.} 
\label{fig:qv-e}
\end{center}
\end{figure}

%-----------------------------------------------------------------------
%
%	Q-INVARIANTS ODD-A ZR
%
%-----------------------------------------------------------------------
\begin{figure}[htb!]
\begin{center}
\includegraphics[width=\linewidth]{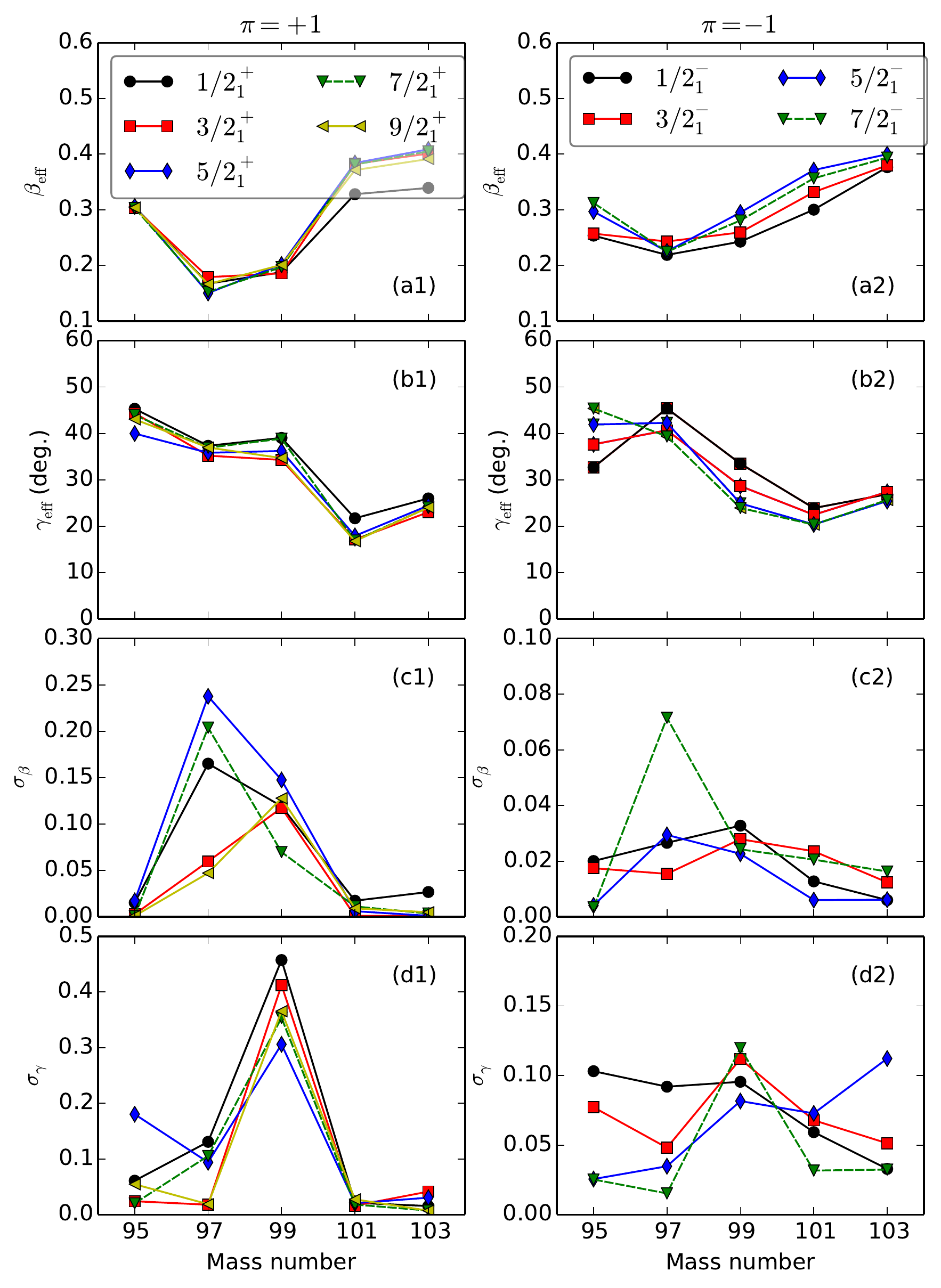} 
\caption{(Color online) 
Same as in the caption to Fig.~\ref{fig:qv-e}, but for several low-spin yrast states 
with positive (left column) and negative (right column) 
parity in the odd-A Zr isotopes. } 
\label{fig:qv-o}
\end{center}
\end{figure}

\section{Signatures of quantum shape phase transition} \label{sec:qpt}

As a signature of quantum phase transition, 
we consider quadrupole shape invariants \cite{cline1986} 
%(referred to hereafter as q-invariants) 
computed using the IBM-2 and IBFM-2 wave functions. 
The relevant  quadrupole shape invariants 
for a given IBM-2/IBFM-2 state $\ket{\alpha I}$, 
where a label $\alpha$ distinguishes states with the 
same spin $I$, are defined as \cite{werner2000}
\begin{align}
&q_2
=\bra{\alpha I}\bigl(\hat Q\cdot\hat Q\bigr)\ket{\alpha I}
\label{eq:q2}
\\
&q_3
=-\sqrt{\frac{35}{2}}\braket{\alpha I | [\hat Q\hat Q\hat Q]^{(0)} |\alpha I}
\label{eq:q3}
\\
&q_4
=\braket{\alpha I | \bigl(\hat Q\cdot\hat Q\bigr)\bigl(\hat Q\cdot\hat Q\bigr) |\alpha I}
\label{eq:q4}
\\
&q_6
=\frac{35}{2}\braket{\alpha | [\hat Q\hat Q\hat Q]^{(0)}[\hat Q\hat Q\hat Q]^{(0)} |\alpha I}
\label{eq:q6}
\end{align}
where 
$[\hat Q\hat Q\hat Q]^{(0)}=[[\hat Q\times\hat Q]^{(2)}\times\hat Q]^{(0)}$, and $\hat Q$ is the 
corresponding E2 transition operator. 
%The explicit forms of the q-invariants in terms of the the E2 reduced matrix elements 
%and their derivations are found in the literature, e.g., in \cite{werner2000}. 
The following dimensionless parameters read:  
$K_n = q_n/q_2^{n/2}$ with $n=3, 4$, and 6, provide the link to the usual deformation parameters 
that characterize the shape of a nucleus:
\begin{align}
& K_3 = \frac{\braket{\beta^3\cos{3\gamma}}}{\braket{\beta^2}^{3/2}}\equiv \cos{3\gamma_\mathrm{eff}} \\
& K_4 = \frac{\braket{\beta^4}}{\braket{\beta^2}^2} \\
& K_6 = \frac{\braket{\beta^6\cos^2{3\gamma}}}{\braket{\beta^2}^3}.
\end{align}
The effective quadrupole deformation parameters read  
\begin{align}
\label{eq:beta}
& \beta_\mathrm{eff} = \sqrt{\braket{\beta^2}} = \frac{4\pi}{3eZR^2}\sqrt{q_2} \\
& \gamma_\mathrm{eff}=\frac{1}{3}\arccos{K_3},
\end{align}
and the corresponding fluctuations of $\beta$ and $\cos{3\gamma}$ can be computed from
%which were considered in Ref.~\cite{werner2000}, as follows
\begin{align}
& \sigma_\beta
= 
\frac{\braket{\beta^4}-\braket{\beta^2}^2}{\braket{\beta^2}^2} 
= K_4 - 1
\\
\label{eq:sig-gam}
& \sigma_\gamma
= 
\frac{
\braket{\beta^6\cos^2{3\gamma}}
- \braket{\beta^3\cos{3\gamma}}^2
}
{\braket{\beta^2}^3}
= K_6 - K_3^2\;.
\end{align}
Note that $R=1.2A^{1/3}$ fm in Eq.~(\ref{eq:beta}). 

In Figs.~\ref{fig:qv-e} and \ref{fig:qv-o} we display 
$\beta_\mathrm{eff}$, $\gamma_\mathrm{eff}$, 
$\sigma_\beta$, and $\sigma_\gamma$, for the even-even and 
odd-A Zr nuclei, respectively. 
The signature of a quantum phase transition can be 
identified as an abrupt change of an 
order parameter for a particular value of the control parameter. In the present case, 
in which we consider geometric shape transitions along a chain of isotopes, 
the neutron number plays the role of the control parameter, while shape invariants or 
effective (state-dependent) deformations can be considered as order parameters.  
The quantities defined in Eqs.~(\ref{eq:beta}) to (\ref{eq:sig-gam}), that is, the 
effective deformations and corresponding fluctuations for the lowest lying states,  
display discontinuities close to the transitional nucleus $^{98}$Zr, at which even-even systems 
undergo a phase transition. 
For the even-even isotopes, in Fig.~\ref{fig:qv-e}, the 
effective deformations $\beta_\mathrm{eff}$ of the lowest three $0^+$ states 
increase smoothly with the neutron number and, as a consequence, 
the fluctuation $\sigma_\beta$ does not change much in the vicinity of $^{98}$Zr. 
The particularly large $\sigma_\beta$ at $^{94}$Zr 
indicates significant shape mixing. 
The effective $\gamma$ deformation, however, exhibits a more pronounced  
change in the transition from $^{96}$Zr to $^{98}$Zr for all three 
$0^+$ states. We note, in particular, the large fluctuations in $\gamma$ 
for the second $0^+$ state.

As shown in Fig.~\ref{fig:qv-o} for the odd-A Zr nuclei, the effective deformations and corresponding fluctuations of 
the lowest positive- and negative-parity states exhibit discontinuities characteristic of a shape phase transition 
at $^{99}$Zr. It is interesting to note that the sudden changes appear to be more pronounced than in the 
even-even neighbors. A similar effect has been found in the analysis of the microscopic signatures of nuclear 
ground-state shape-phase transitions in odd-mass Eu isotopes \cite{quan2018}, and attributed to a shape polarization 
effect of the unpaired nucleon. 
In the present case the strongest signature of a shape phase transition is provided by 
the effective deformations and their fluctuations for the lowest positive-parity states. Pronounced discontinuities  
appear between $^{99}$Zr and $^{101}$Zr, and their microscopic origin can be clearly identified in the composition of 
the IBFM-2 wave functions shown in Fig.~\ref{fig:wf}.
We note that the enhancement of a shape phase transition 
in the presence of an unpaired nucleon has also been explored using a more 
phenomenological IBFM approach \cite{petrellis2011}.

\section{Summary\label{sec:summary}}

Spectroscopic properties relevant for the characterization of shape phase transitions  
in even-even and odd-A neutron-rich Zr isotopes have been investigated using 
the microscopic framework of nuclear DFT. Deformation 
constrained SCMF calculations have been performed with the
relativistic Hartree-Bogoliubov method based on the universal energy 
density functional DD-PC1 and a separable pairing interaction. 
The triaxial $(\beta,\gamma)$ deformation energy surfaces obtained 
from the SCMF calculations for the 
even-even $^{94-102}$Zr isotopes predict a very interesting nuclear 
structure evolution: shallow triaxial deformations in $^{94}$Zr, 
a $\gamma$-unstable potential in $^{96}$Zr, coexistence of 
a shallow oblate and strongly-deformed prolate minimum 
in $^{98}$Zr, and the occurrence of $\gamma$-softness in $^{100,102}$Zr. 
These SCMF results corroborate the 
conclusions of recent experimental studies.

The excitation spectra of the even-even Zr nuclei have been computed 
by mapping the SCMF deformation energy surfaces onto the expectation 
value of the IBM-2 Hamiltonian in the boson condensate state. 
A phase-transitional behavior of the low-lying excitation spectra, that occurs between 
$^{96}$Zr and $^{100}$Zr, is qualitatively reproduced. 
The excitation energies of the low-lying second $0^+$ in $^{98,100}$Zr 
are, however, considerably overestimated in the present calculation. 
These low-lying $0^+$ excitation energies have previously been 
explained by effects such as shape coexistence related to intruder configurations or pairing vibrations, 
both of which are outside the configuration space of the present IBM framework.

Spectroscopic properties of the odd-A Zr 
nuclei are computed by means of the particle-core coupling 
of the IBFM. The SCMF calculations provide a microscopic input 
for the construction of the basic parts of the IBFM Hamiltonian. 
The calculated low-energy spectra 
of the odd-A Zr isotopes exhibit interesting structural evolution 
close to the neutron number $N=59$, and are in  
very good agreement with the experimental results. 
In $^{95,97}$Zr, both the positive- and negative-parity spectra correspond  
to a weak coupling of a vibrational even-even core to the odd 
particle (neutron in this case). For $^{101,103}$Zr, bands typical of the odd nucleon 
strongly coupled to a well-deformed even-even core appear as yrast structures. 
The low-energy spectra for the transitional nucleus $^{99}$Zr 
can be characterized by the coexistence of $\Delta I=1$ and 
$\Delta I=2$ positive-parity bands.  
The calculated quadrupole shape invariants provide a 
signature of a shape phase transition. 
The interesting result is that, for the 
odd-A Zr isotopes, the effective deformations $\beta$ and 
$\gamma$, and their fluctuations exhibit more pronounced discontinuities   
at the point of shape phase transition when compared to their 
even-even neighbors. 

Taking into account that a microscopic SCMF calculation based 
on a universal EDF completely determines 
the even-even core Hamiltonian and most of the 
IBFM Hamiltonian, and that only a few adjustable 
parameters specify the fermion-boson terms, 
this approach holds promise for exploring 
simultaneously even-even and odd-mass neutron-rich 
nuclei in this challenging region of the nuclear chart. 
A prospect for future studies is to improve the description 
of the even-even Zr nuclei, especially the low-lying excited $0^+$ states. 
In this respect, a configuration-mixing IBM calculation 
based on the Gogny HFB has already been reported for 
the even-even Zr isotopes \cite{nomura2016zr}. 
It will be interesting to develop a formalism that incorporates these 
additional effects consistently both for even-even 
and odd-A systems.

\begin{acknowledgments}
This work has been supported by the Tenure Track Pilot Programme of 
the Croatian Science Foundation and the 
\'Ecole Polytechnique F\'ed\'erale de Lausanne, and 
the Project TTP-2018-07-3554 Exotic Nuclear Structure and Dynamics, 
with funds of the Croatian-Swiss Research Programme. 
It has also been supported in part by the QuantiXLie Centre of Excellence, a project co-financed by the Croatian Government and European Union through the European Regional Development Fund - the Competitiveness and Cohesion Operational Programme (KK.01.1.1.01).
\end{acknowledgments}

\bibliography{refs}

\end{document}